SECTION I

# Metrics

CHAPTER 1

# Usage Bibliometrics

*Michael J. Kurtz*
*Harvard-Smithsonian Center for Astrophysics, Cambridge, MA*

*Johan Bollen*
*Los Alamos National Laboratory, Los Alamos, NM*

## Introduction

How researchers access and read their technical literature has gone through a revolutionary change. Whereas fifteen years ago nearly all use was mediated by a paper copy, today nearly all use is mediated by an electronic copy. Within individual disciplines the change has been nearly instantaneous. As an example, in mid-1997 the number of papers downloaded from astronomy's digital library, the Smithsonian/NASA Astrophysics Data System (ADS; ads.harvard.edu) exceeded the sum of all the papers read in all of astronomy's print libraries, combined (Accomazzi, Eichhorn, Kurtz, Grant, & Murray, 2000; Eichhorn, Kurtz, Accomazzi, Grant, & Murray, 2000; Grant, Accomazzi, Eichhorn, Kurtz, & Murray, 2000; Kurtz, Karakashian, Grant, Eichhorn, Murray, Watson, et al., 1993; Kurtz, Eichhorn, Accomazzi, Grant, Demleitner, & Murray, 2005; Kurtz, Eichhorn, Accomazzi, Grant, Murray, & Watson, 2000). This was six months after two of astronomy's four largest journals became available electronically, and six months before the other two were available online.[1] Studies (Tenopir, King, Boyce, Grayson, & Paulson, 2005; Kurtz, Eichhorn, Accomazzi, Grant, Demleitner, & Murray, 2005) show that the ADS is used on a near-daily basis by a large majority of research astronomers. This rapid transformation took place across all scientific domains; in November 2006, Elsevier reported one billion downloads since the inception of its ScienceDirect service in 1999, a number that greatly exceeds the total number of citations published since the 1900s (www.info.sciencedirect.com/news/press_releases/archive/archive2006/one_billionths.asp).

A key difference between print and electronic libraries is the detailed records that electronic libraries keep of every transaction. In a paper library a typical researcher might browse a book of abstracts, find some articles of interest, retrieve them from the stacks, photocopy them, and return to her office to read them, leaving no trace, other, perhaps, than that captured by reshelving statistics. Today she would accomplish this with a few commands to a search engine and a few clicks of the mouse, all from her desk, and all such actions would be recorded. Indeed, today, for every single use of an electronic resource, the system can record which resource was used, who used it, where that person was, when it was used, what type of request was issued, what type of record it was, and from where the article was used. Although these data do not include essentially unknowable quantities such as user motivations or goals, they do constitute a much more comprehensive record of usage than was previously available. When, in addition, these data are merged with bibliographic data, such as authors and citations, we have a very accurate record of user activity.

Methods have been and will be created to measure the flow of information among countries, disciplines, and groups of individuals, and to assess the productivity of a number of different entities on the basis of citations, number of publications, and other text-based data. These methods play an increasingly important role in the allocation of resources, but their present focus on citation and text data entails a number of distinct disadvantages. Citation data concern primarily journal articles and their authors, are subject to significant publication delays, and offer one particular perspective on scholarly activity that overlooks the activities of those not associated with the present publishing (and citation) system.

Large-scale usage data on the other hand are not subject to publication delays, not limited to journal articles and their authors, and offer a fairly comprehensive overview of activities within all phases and social layers of the scientific process. It is thus inevitable that this wealth of data will be increasingly integrated in bibliometrics. Indeed, following the success of usage data analysis in commercial environments, such as recommender systems employed by Amazon.com and Netflix.com, the past five years have seen a surge in bibliometric investigations of usage. The introduction of this new usage data environment is, however, both extremely powerful and extremely dangerous; it adds significant capabilities to the instrumentarium of bibliometric analysis, but does so on the basis of data whose characteristics and validity have only recently become the focus of scientific investigation.

The growing role of usage data necessitates an expanded definition of bibliometrics. Bibliometrics is generally defined (Broadus, 1987) as the quantitative study of published units on the basis of citation and text analysis, but can include studies based on usage data. Small-scale, institutionally bound studies of reshelving statistics have indeed played a minor role in traditional bibliometrics. However, this type of usage data is qualitatively and quantitatively different from that collected by modern electronic library services. Whereas well-vetted, large-scale databases have been available for bibliometric analysis for several decades, this has not been the case for usage data. Their limited scale, detail, and applicability have imposed strict limitations on their role in traditional bibliometrics.

In this chapter we are concerned with an expanded definition of bibliometrics that includes modern usage data that approximate or even surpass the scale, quality, and detail of citation and text databases. These usage data: are extremely large-scale; are sample significant and diverse portions of the scholarly community; are highly detailed in their request-metadata; allow reconstruction of user clickstreams; are recorded at the article level; and are now commonly recorded by the electronic services offered by publishers, aggregators, and institutional repositories. Present developments indicate an expanded role for this type of usage data that will revolutionize bibliometrics as we presently know it.

By our definition user-centric studies would come under the broader classification of user-based (scholarly) informetrics (Bar-Ilan, 2008; Wilson, 1999) and are beyond the scope of this review. For discussions of user behavior, the interested reader might consult the recent reviews by Jamali, Nicholas, and Huntington (2005), King and Tenopir (1999), Rowlands (2007), and Wang (1999). For discussions of searching behavior we suggest the papers by Borgman, Hirsh, and Hiller (1996), Hider, (2006), and Jansen (2005). Outside the realm of scholarly literature, user studies are an important component in the commercial world. Marketing, advertising, and product design are some obvious examples. In addition, the recent compendium "Usage Statistics of E-Serials" edited by Fowler (2007) contains 17 articles addressing various aspects of usage studies; this volume concentrates mainly on practical issues of immediate interest to practicing librarians; the present review is more concerned with theoretical and longer term issues; of particular interest here is the review of the MAXDATA project by Tenopir, Baker, Read, Manoff, McClanahan, Nicholas, and colleagues (2007). Luther (2000) reviewed the field a decade ago from the point of view of librarians.

This expanded notion of bibliometrics includes two general clusters of study. First, one focused on describing and modeling individual user behavior, for example, to improve user interfaces and to study user motivations. The second is to a greater degree concerned with the actors and units involved in the production of scholarly artifacts themselves, for example, ranking an article or an author as a function of usage. The latter does not require individual user identification and is therefore less affected by concerns of user privacy or user rights. We will not discuss these issues at length, but we note that user privacy remains an issue for all user studies; data sharing between researchers is a particular concern.

In this review we are principally concerned with how usage data may be defined and collected; we discuss some of the ways it has been used and combined with traditional bibliometric data; we also discuss some problems involved in collecting the data and obtaining fair or representative samples. Due to the exigencies of currently available data we restrict our discussion to (online) journal articles, leaving to another time the newer forms of scholarly communication, such as data archives, online databases, work flow systems, and blogs.

## Usage Data and Statistics

A review of the literature on usage-based bibliometrics shows that the notion of usage has been defined and operationalized in a variety of ways. Some authors have discussed usage in terms of "reads"; others have preferred "uses," "downloads," or "hits." A formal definition of usage must acknowledge the variety of contexts in which usage-based bibliometrics can take place but should not include unmeasurable quantities such as user motivations and intentions or context-specific issues such as interfaces and system configurations.

### *A Request-Based Model of Usage*

Usage data can be and have been recorded in a variety of contexts. However, when we examine the actors and entities that are typically involved, we find a number of common elements. First, we have the *user* who may or may not be personally identifiable but can be assumed to have an *interest or need* with regard to a particular resource. Second, we have the *information service*, which functions as a mediator between the *user* and the set of *resources*. Third, we have the *scholarly resources* themselves, which include books, journal articles, and e-science data sets.

When an agent has a particular interest or need for a particular resource, he or she issues a *request* to the information service. The information service processes the request and returns a *service* pertaining to the resource or some representation of the resource. The information service will have a vocabulary of requests to which it will respond and a set of services it can render. An overview of this model is shown in Figure 1.1.

We can now define the notion of usage in terms of these transactions:

> *Definition 1*: Usage occurs when a user issues a request for a service pertaining to a particular scholarly resource to a particular information service.

And consequently the notions of usage event and usage data:

> *Definition 2*: A usage event is the electronic record of a user-generated request for a particular resource, mediated by a particular information service, at a particular point in time. Usage log data are collections of individual usage events recorded for a given period of time.

Such requests generally result from some degree of user interest in the resource. However, they can serve as only a post hoc operationalization of user motivations, interests, needs, intentions, or post-request usage that led to the request. It is also assumed that different request types can express various degrees of user interest, for example, a request to view an article's abstract represents a lower degree of user interest in the article than a request for a full-text download. The relation between user motivations and requests, and the relation between the degree of interest to particular request types, are the subject of ongoing research.

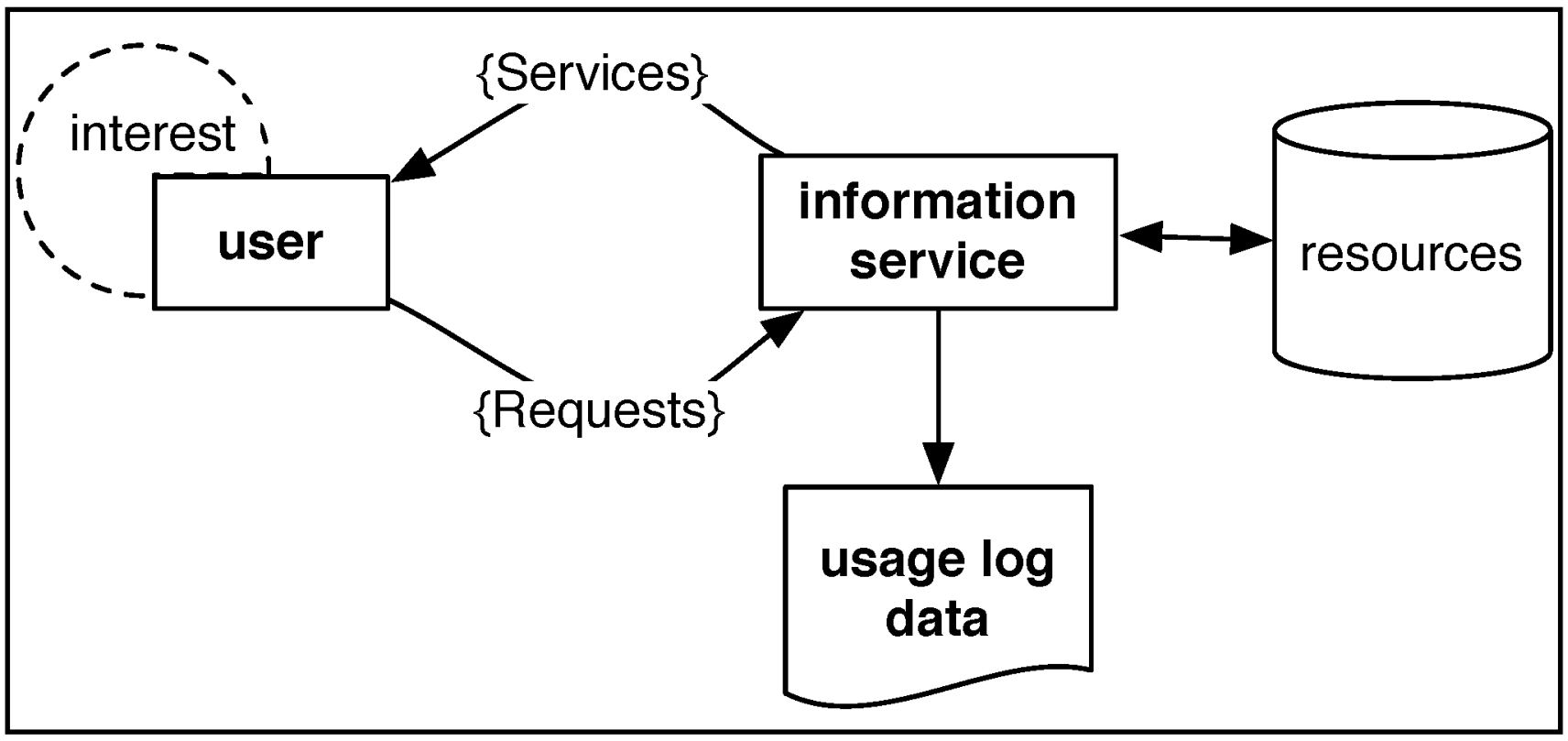

**Figure 1.1 Service request model underlying definition of usage**

Usage data can be recorded at any point in the processing of the request, but it is most commonly recorded by the information service at the time the request is fulfilled. This leads to usage data that generally capture the following information:

- *Event identifier* distinguishes individual events.
- *User or session identifier* distinguishes individual users or user sessions.
- *Request type identifier* indicates the type of user request issued, e.g., full-text download, abstract view, etc.
- *Resource identifier* or *resource metadata* provides a unique identifier of the resource for which the service was requested, or sufficient metadata (title, author name, etc.) to identify it.
- *Date and time of the request.*

We shall discuss how these information fields, although minimal in comparison with what is generally recorded, still provide sufficient information to support sophisticated bibliometric analysis.

## Usage Data versus Usage Statistics

It is quite common for usage data to be aggregated into usage statistics by an aggregation function. This function can be defined to perform any desired aggregation, for example, counting the number of events pertaining to a resource within a particular period of time. The result of this aggregation is referred to as a *usage statistic*. This process can be formally represented as follows:

$$S = \Sigma(U, R, P)$$

where $S$ is the usage statistic resulting from applying the aggregation function $\Sigma$ to the usage data $U$ with regard to resource $R$ using the set of parameters $P$ that controls the aggregation process, for example, a date range $d \in P$, $d = [d_1, d_2]$ so that $\Sigma$ is restricted to usage occurring at time $d_1 < t < d_2$ and $S$ will reflect cumulative usage rates for that period.

In contrast to usage data, usage statistics are generally characterized by the absence of all individual event information. However, the resource identifier $R$ will generally be retained because it is the referent of the usage statistic, in addition to the value of the usage statistic $S$, and possibly a period of time for which the aggregation was conducted as part of the parameter set $P$. Table 1.1 provides an overview of the main differences between usage data and usage statistics.

**Table 1.1 Comparison of features generally present or lacking in usage data and usage statistics**

| field | data | statistics |
|---|---|---|
| event ID | Yes | No |
| user ID | Yes | No |
| session ID | Yes | No |
| request type | Yes | Yes |
| resource ID | Yes | Yes |
| date-time | Yes | Yes |
| aggregate value | No | Yes |

Many usage-derived indicators of impact or status can be expressed as instances of the aggregation function $\Sigma$ wherein the parameter set $P$ is defined to yield a normalized or unnormalized impact metric over a particular period of time. The proposed Usage Impact Factor (Bollen &

Van de Sompel, 2008) or Usage Factor (Shepherd, 2007), defined as the average usage rate of a journal's articles over a two-year period, is an example of such a metric derived from aggregate usage statistics. The Metrics from Scholarly Usage of Resources (MESUR) project's (www.mesur.org) ontology (Rodriguez, Bollen, & Van de Sompel, 2007) represents an attempt to model the various relationships among resources, users, aggregation functions, and other metadata associated with such usage indicators.

## *Usage Data: A Practical Overview*

### Usage Data in the Pre-Electronic Era

Past technological limitations limited the recording of usage data to on-site library usage of printed matter. For example, Scales (1976), Galvin and Kent (1977), and more recently King, Tenopir, and Clarke (2006) operationalize usage in terms of reshelving and circulation statistics. Similarly, Tsay (1998b) determines journal usage from reshelving rates and has found statistically significant correlations between such journal usage and citation impact rankings.

We can frame these studies in terms of the request-based model. In this case a usage event is defined as the record of a user issuing a request to the information services of a physical library. The request can consist of physically retrieving a journal or book from the library's shelves, requesting that a hold be placed on a particular item, using interlibrary loan, or simply checking out the item. The fulfillment of the request consists of the physical transfer of the item from the library's collection to the user. The request and service are thus closely entangled because the physical library acts as the service provider (mediator) and as a collection of resources.

This yields usage data that share many features with usage statistics as listed in Table 1.1.

- *Absence of context*: Due to the physical entanglement of service and collection, little information can be recorded with regard to the context in which the request takes place, for example, types of user request, exact date and time of the request, and, therefore, the sequence or temporal relation of subsequent requests.
- *Loss of resource information*: Circulation and reshelving statistics generally do not distinguish between individual articles but are recorded at the journal and book level.
- *Scale*: Reshelving and circulation statistics are limited to particular library systems and their patrons.

This limits the role of circulation and reshelving statistics to that played by usage statistics, by contrast with usage data, which constitute a more detailed, contextual record of individual usage events.

## Usage Data from Web Server Logs

Since the early 1990s an increasing number of libraries and information services started to fulfill user requests via their Web servers. The resulting Web server "logs" record the parameters of the HTTP requests that were issued by users in their interactions with the particular service. This log thus reflects usage from the perspective of a Web server functioning as the mediator between users and the information service.

Given the prevalence of the Apache HTTP server, its Common Log Format (CLF) (httpd.apache.org/docs/logs.html) has become a *de facto* standard for usage data, shaping the log data recorded by other HTTP servers as well. The Apache CLF stores a number of fields pertaining to individual HTTP requests, such as the client's IP address, a user ID (if determined by HTTP authentication), the date and time at which the server processed the request, the actual HTTP request issued by the client, the request status code indicating whether the request was successfully fulfilled, and finally the byte size of the object returned to the client. In addition, it is possible to store the "Referrer," which corresponds to the site from which the client reports to have been referred. As such the Apache CLF conforms to the minimal requirements specified in our request-based model of usage.

A particular issue is the absence of reliable session information, which hampers the reconstruction of the sequences of individual user requests. Because many users do not have permanent IP addresses and access library services using proxies, the IP address does not adequately identify individual users or sequences of requests issued by the same user. The latter are crucially important to capture how users move from one resource to another in their usage "clickstreams." These are vital to modeling user traffic, determining resource relationships, and implementing personalized services. In addition, advanced social network impact indicators can be derived from resource relationships extracted from user clickstream data. The statistical reconstruction of session information from Web server logs is therefore a matter of considerable interest (He, Goker, & Harper, 2002; Heer & Chi, 2002; Pirolli & Pitkow, 1999).

Although the Apache CLF is frequently used to record usage data, the usage it records may be modulated by middleware applications such as EZproxy (www.oclc.org/ezproxy/about/default.htm) and other customized and ad hoc environments that control access and authentication. In such cases, the HTTP requests may contain additional information, for example, user identification code and session data, encoded in the HTTP request or request URL. In addition, EZproxy itself stores a record of user accesses that is recorded in a standard Web server log file format.

## Link Resolvers

In the past five years scholarly information services have started to support context-sensitive services (Van de Sompel & Beit-Arie, 2001a,

2001b) by implementing the OpenURL 0.1 specification. This development has been met by the widespread installation of linking servers by academic and research libraries to provide localized services. Link resolvers serve as a hub in the institutional information environment; they can therefore record usage across a variety of different scholarly information services.

Figure 1.2 provides an example of the pivotal role link resolvers play in allowing institutions to record their communities' usage across various information services. Each of the OpenURL-enabled information services, for example, Google Scholar, inserts an OpenURL into every reference to a scholarly work that is presented to a user. The OpenURL consists of an HTTP GET request to the institutional link resolver that contains metadata to identify the referenced work. Upon receiving the OpenURL, the institutional link resolver can offer a list of customized services pertaining to the referenced work, typically those of other information services that are available in the user's distributed information environment, such as full-text database systems.

Because user selections across a variety of OpenURL-enabled services are routed back to the institutional link resolver, it can track user requests across various OpenURL-enabled services. The ability to collect institutional usage data that spans across many OpenURL-enabled information services and the reliance on a common standard to represent usage data have increased use of link-resolver-generated usage data in bibliometric research, for example, McDonald (2007) and Bollen and Van de Sompel (2008).

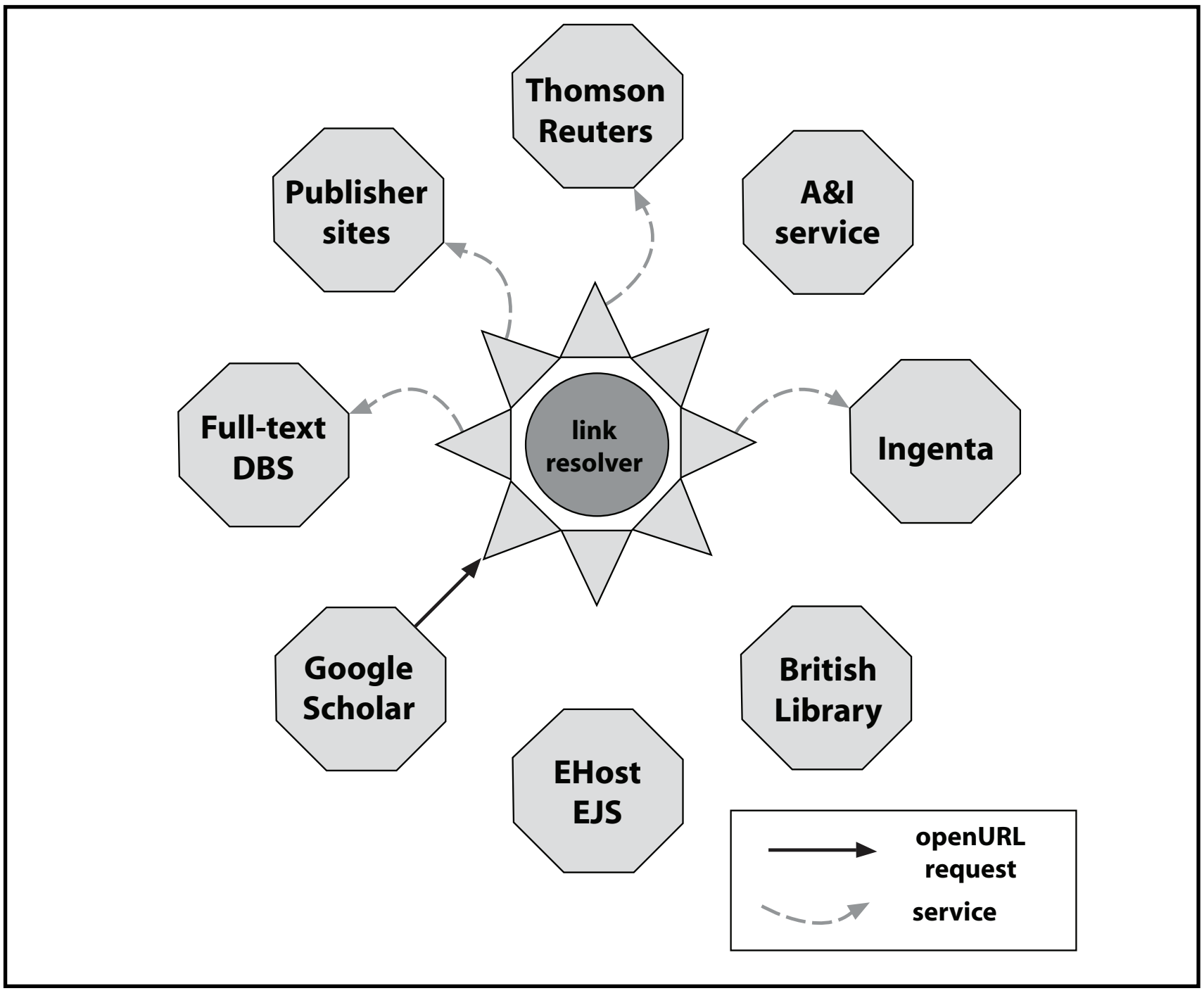

**Figure 1.2 Linking servers are well positioned to capture usage data.**

In addition to the unique capability to track user requests across different scholarly information services, link resolvers can rely on a common representation framework for usage data as provided by the ISO/ANSI Z39.88-2004 standard "The OpenURL Framework for Context-Sensitive Services" (library.caltech.edu/openurl/default.htm). The Z39.88-2004 standard rests on the notion of a ContextObject (shown in Figure 1.3) whose structure contains six different fields required to fulfill user requests and thus offers a representational framework for usage data as defined in our request-based model of usage.

- *Referent*: Subject of the service request that the ContextObject encodes
- *Requester*: Agent that requests the service pertaining to the *Referent*
- *ServiceType*: Type of service that is requested
- *ReferringEntity*: Entity that references the *Referent*
- *Resolver*: Target of a service request
- *Referrer*: The system that generated the ContextObject

The OpenURL ContextObject can as such express the relevant features of most usage events. Bollen and Van de Sompel (2006a) propose an eXtensible Markup Language (XML) serialization of the ContextObject in which the date-time stamp of the event is stored in the administrative element of the XML Context Object, namely the timestamp attribute, as well as a globally unique event identifier in the form of a Universally Unique Identifier (UUID). The resulting usage data can then be exposed by Open Archives Initiative Protocol for Metadata Harvesting (OAI-PMH) repositories (Van de Sompel, Young, & Hickey, 2003) and harvested to form usage data aggregations across multiple institutional usage data repositories. Bollen and Van de Sompel (2006a) thus propose using linking servers as *intra-institutional* aggregators of usage information whereas the *inter-institutional* aggregation of usage data takes place by the OAI-PMH harvesting and aggregation of this data to a central location. The proposed system can create usage data aggregations with a global or regional reach.

A number of disadvantages of this approach have to be noted as well. First, linking servers cannot record user requests that do not pass through their services. No linking server usage logs thus will be available in cases where an institution does not employ a linking server or a user visits a service that is not OpenURL-enabled. Second, any fields or parameters associated with usage that are not captured by the OpenURL ContextObject will not be recorded. In other words, the standardized format of the OpenURL ContextObject limits the level of detail with which usage can be recorded. Third, although quite common, the use of linking servers is not universal. Many commercial products exist but may be outside the reach or requirements of many institutional library services. The architecture proposed by Bollen and Van de Sompel (2006a) to aggregate usage data recorded by multiple link resolvers cannot address this issue.

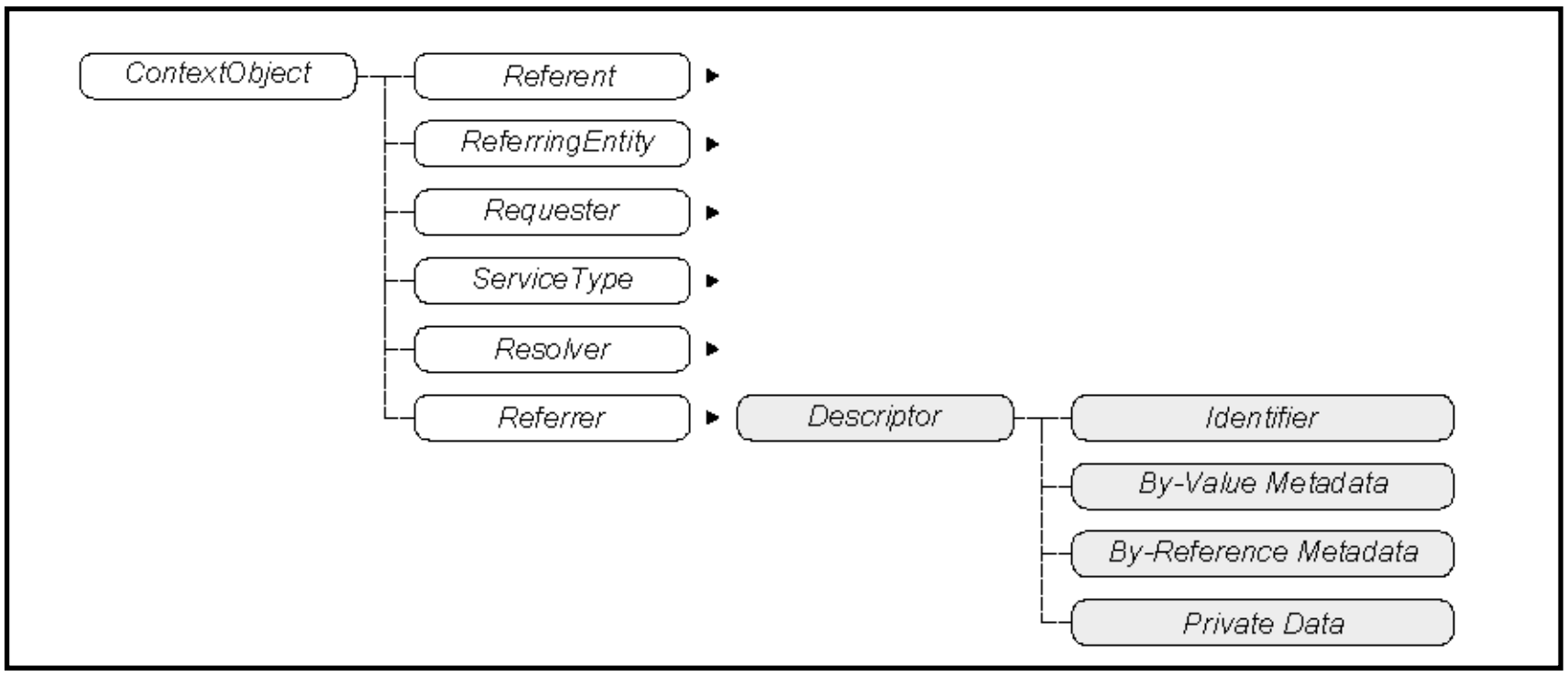

**Figure 1.3 Components of the OpenURL ContextObject (in Bekaert et al., 2005)**

## Usage Statistics: Emerging Standards

The proposal by Bollen and Van de Sompel (2006a) concerns the construction of a standards-based infrastructure to aggregate the usage data recorded by link resolvers across multiple, different communities. In this sense it represents an attempt to standardize the recording, representation, and sharing of usage data.

The Metrics from Scholarly Usage of Resources (MESUR) project has continued this development. The project aims to expand the toolkit for scholarly evaluation by investigating the feasibility of numerous usage-based impact metrics calculated on the basis of large-scale usage data that were aggregated from some of the world's most significant publishers, aggregators, and institutional consortia. Its usage database now stands at approximately 1 billion ($1 \times 10^9$) usage events. Given the variety of formats and recording conditions across MESUR's providers of usage data, the project proposed an RDF/OWL ontology (Rodriguez et al., 2007) that formally describes the semantic relationships among various entities associated with citation and usage data.

Beyond that, very few efforts have focused on architectures and standards recording and aggregating usage data across institutional boundaries. However, considerable progress has been made with regard to the standardization and aggregation of usage *statistics*. In particular, the COUNTER initiative (Counting Online Usage of NeTworked Electronic Resources; www.projectcounter.org) has made significant inroads in recording and publishing *publisher*-generated usage statistics.

The COUNTER initiative has issued Codes of Practice that define standards and protocols with regard to the recording and sharing of vendor-generated usage statistics. Vendors can choose to comply with the COUNTER requirements that define how article and database usage is to be recorded and shared. A mechanism exists to audit the recording process and determine whether compliance has been achieved and maintained. Compliant vendors can issue various types of COUNTER reports that outline, for example, the "number of successful full-text article downloads" (Journal Reports 1 or JR1) or "Total Searches and Sessions by Month and Database" (Database 1 or DB1). COUNTER reports can be issued in a CSV-formatted document that includes the *journal title*, *publisher*, *platform*, *print and electronic journal ISSN*, *monthly totals,* and *year-to-date totals*. Furthermore, the Standardized Usage Statistics Harvesting Initiative (SUSHI; www.niso.org/workrooms/sushi) has defined a standard for the harvesting of electronic resource usage data that allows the exchange of XML-formatted COUNTER-compliant usage statistics.

A number of important distinctions should be noted between COUNTER-compliant usage statistics and usage data as described in our request-based model of usage: The former are journal-level, aggregate, monthly usage *statistics* whereas the latter contain the actual record of each unique user request, including the resources to which the request pertained, the time at which it occurred, and the user session of which it was a part.

## Usage Bibliometrics

Many bibliometric measures based on citations have direct analogs with usage. A citation is an action that refers to a document, as are all types of usage discussed here. Thus a paper, or group of papers (representing, say, an author) can have a citation rate and a usage rate (the number of citations or usage events per unit of time; other rates, such as the number of citations per usage event have also been studied), and these may be combined or analyzed in various ways. In addition to citation and usage rates, networks of related articles and journals can be extracted from citation and usage data and analyzed using methods developed in social network analysis.

Usage obviously has properties different from citation or article counts; it may be expected that measures derived from usage will have both similarities with and differences from these more traditional measures. As a first task it is necessary to demonstrate these similarities and differences, thus beginning to validate the domain where usage-based measures will be important.

## *Direct Measures*

What is the relation between an article's citation rate and its usage rate? Citations are normally referred to by their total, the integral of the rate over the time since publication; usage, although it is theoretically possible to measure it as a total integral over time, is normally measured as an approximately instantaneous usage rate. Systematics, such as the growth of the Internet and the development of new communication techniques—digital libraries, open access, mega-publishers—make the use of time-integrated total usage information problematic.

### Obsolescence

A convenient measure for comparing citations with usage is the obsolescence of articles—the description of how the passage of time affects the citation rate or usage rate for articles or aggregations of articles. Obsolescence based on citation and circulation information has a large literature, beginning with Gross and Gross (1927) and especially Gosnell (1944); White and McCain (1989) review it and Egghe and Rousseau (2000) discuss the mathematical issues. Obsolescence based on electronic usage has a much smaller literature; the first paper extensively exploring its properties and comparing them with citations was by Kurtz and colleagues (2000), which they later expanded substantially (Kurtz, Eichorn, Accomazzi, Grant, Demleitner, Murray, et al., 2005).

Since then, Moed (2005) discussed the relation between usage and citations for articles within a single journal; Ladwig and Sommese (2005) tried to adjust usage statistics by using citation obsolescence measures to account for systematic differences in collected usage measures; the University College London (UCL) group studied usage obsolescence using the transaction logs of publishers (Nicholas, Huntington, Dobrowolski, Rowlands, Jamali, & Polydoratou, 2005) and using the transaction logs of a library consortium (Huntington, Nicholas, Jamali, & Tenopir, 2006); Duy and Vaughan (2006) looked into the possibility of replacing citation measures with usage measures; and McDonald (2007) used the logs of the CalTech library to examine the relation between usage and citations following the introduction of an OpenURL server (this paper also has an extensive bibliography). Tonta and Unal (2005) looked at use obsolescence via document delivery requests.

Figure 1.4, from the paper by Kurtz, Eichhorn, Accomazzi, Grant, Demleitner, Murray, et al. (2005), shows usage as a function of publication date, in terms of uses per article per year for the three principal U.S. astronomy journals (*Astronomical Journal*, *Astrophysical Journal*, *Publications of the Astronomical Society of the Pacific*); the publication dates range from 1889 to 2000. The data came from the transaction logs of the ADS (Kurtz et al., 1993, 2000; Kurtz, Eichhorn, Accomazzi, Grant,

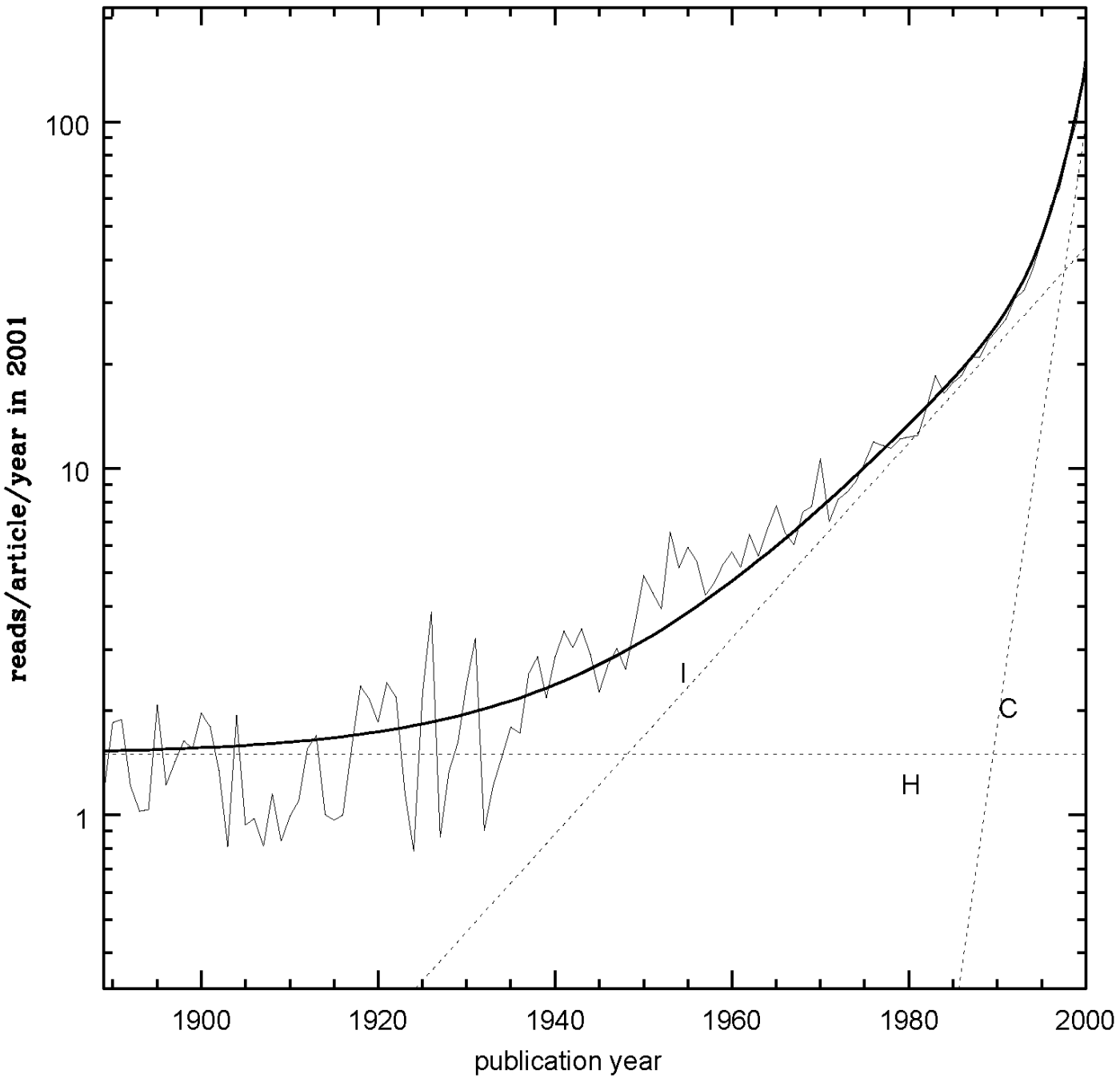

**Figure 1.4 Mean per article use (not differentiated by usage type) in 2001 of the three main U.S. astronomy journals, *Astronomical Journal*, *Astrophysical Journal*, and *Publications of the Astronomical Society of the Pacific*, as a function of publication date (1889–2000), compared with the readership model *R* in equation 1 (smooth solid line) and three of its components (I,C,H) (Kurtz, Eichhorn, Accomazzi, Grant, Demleitner, Murray, et al., 2005). Notice that the *R* model matches the data very well.**

Demleitner, & Murray, 2005) for the first 7.66 months of the year 2001, and represent 1.8 million separate usage events.

These data may be (non-uniquely) modeled as the sum of three exponentials and a constant (Kurtz et al., 2000; Kurtz, Eichhorn, Accomazzi, Grant, Demleitner, Murray, et al., 2005):

$$R(t) = R_H + R_I + R_C + R_N \quad (1)$$

where

$$R_H = H_0 e^{-k_H t}$$
$$R_I = I_0 e^{-k_I t}$$
$$R_C = C_0 e^{-k_C t}$$
$$R_N = N_0 e^{-k_N t}$$

and

$$H_0 = 1.5; k_H = 0$$
$$I_0 = 45; k_I = 0.065$$
$$C_0 = 110; k_C = 0.4$$
$$N_0 = 1600; k_N = 16$$
$$t = \text{time since publication in years}$$

The four functions (three are represented in the figure) may be interpreted as four different modes of user behavior: $N$, New, represents the browse mode use of newly released articles, of which browsing the table of contents of the current issue of a journal would be an example; $C$, Current, represents non-specific searches for current articles, such as searches for recent papers on a subject or by an author; $I$, Interesting, represents searches for specific articles, such as from a reference list of another paper; $H$, Historical, is an approximately constant usage, nearly independent of the age of an article, and could perhaps be called random. There may well be other, more compelling interpretations of the shape of the obsolescence graph, in terms of user behavior, but that is outside the scope of this review.

Figure 1.5, also from the paper by Kurtz, Eichhorn, Accomazzi, Grant, Demleitner, Murray, et al. (2005), shows an expanded view of Figure 1.4 covering 25 years. Comparing the two figures makes clear that the data are well fit by the three components, $C$, $I$, and $H$, and that all three are necessary. The $N$ component of use is too short term to be visible on these graphs; Henneken, Kurtz, Eichhorn, Accomazzi, Grant, Thompson, and colleagues (2006, 2007), using access logs from arXiv (Ginsparg, 1994, 2001; Ginsparg, Houle, Joachims, & Sul, 2004), show the short term obsolescence of physics and astrophysics articles; the $N$ mode is clearly visible in the data from Henneken and colleagues.

Although they, perhaps wisely, do not attempt a parameterization, Huntington and colleagues (2006) confirm the four component usage model presented by Kurtz and colleagues (2000; Kurtz, Eichhorn, Accomazzi, Grant, Demleitner, Murray, et al., 2005). Using logs from a large university consortium (OHIOLINK) they noted the behavior of individuals viewing the table of contents "current awareness checkers" (the $R_N$ mode). They did not explicitly separate the $R_C$ and $R_I$ modes, but they note the characteristic shape of what they called "the sharp decline phase," which "spanned the first 8 to 9 years from publication date" (the $R_I$ mode); "decline was most evident in the first 2 to 3 years" (the $R_C$ mode). They then described the $R_H$ mode: "after the sharp decline period there followed a relatively stable or flat period of usage."

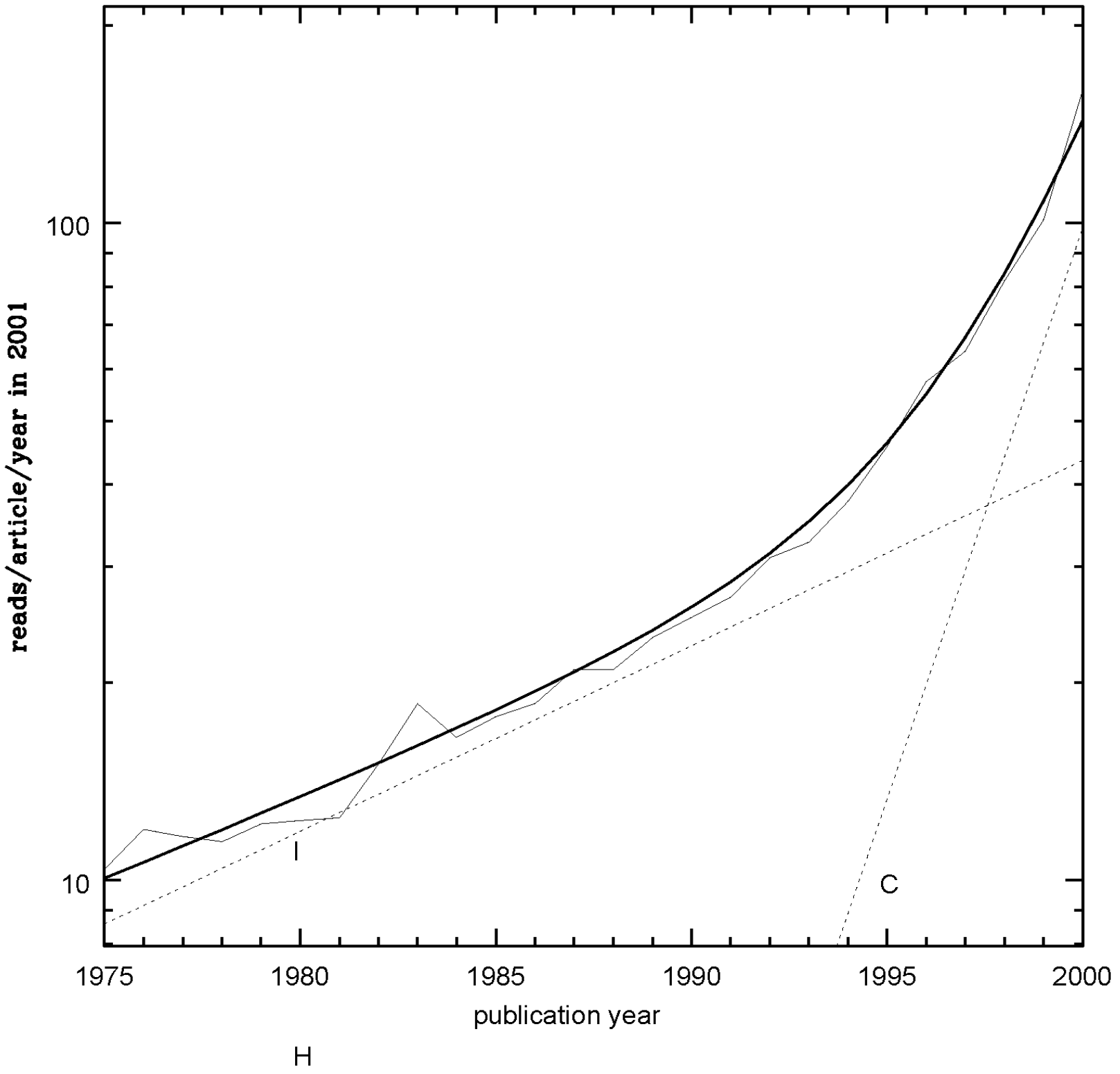

**Figure 1.5 An expansion of Figure 1.4 showing only the most recent 25 years, 1975–2000. Again the *R* model in equation 1 matches the data very well (Kurtz, Eichhorn, Accomazzi, Grant, Demleitner, Murray, et al., 2005).**

Work by Huntington and colleagues (2006) and the companion paper by Nicholas and colleagues (2005), which uses publisher logs, are the two papers from the UCL group that most closely correspond to our broad statistical (bibliometric) viewpoint, as presented in this review. David Nicholas and his very prolific (more than 100 papers) collaborators at UCL have created a large body of work looking at transaction log data similar to that discussed here; their viewpoint is, however, substantially more informetric, or user behavior centric than ours. This independent view stands on its own and deserves separate consideration; Rowlands (2007) and Jamali and colleagues (2005) present recent reviews.

## Sample Effects

Different sets of users have different patterns of use; whether physicists or physicians, astronomers or astrologers, groups have separate literatures and their own unique ways of using them. Bollen and Van de Sompel (2008) examined the usage patterns of an undergraduate university and found a substantial disparity between the actual use and that which might be predicted from the impact factor (Garfield, 1972, 2006). This disparity lessens substantially in areas where use is predominantly by post docs and professors. Huntington and colleagues (2006) found use could have an event-driven component, such as when students receive a reading assignment.

Figure 1.6 shows per article usage of papers from the four main astronomy journals (*Astronomical Journal*, *Astronomy and Astrophysics*, *Astrophysical Journal*, and *Monthly Notices of the Royal Astronomical Society*) as a function of publication date. The data come from the ADS logs for October 2007. The thick line represents the use by heavy (more than 10 queries per month) ADS users, essentially all astronomers; the thin line represents the use by individuals who enter ADS via a link from a Google query. The different sets of users yield very different obsolescence curves. The thick dotted line is the same model curve as in Figures 1.4 and 1.5. Save for the current year, which is strongly effected by $N$ mode behavior due to the introduction of the myADS notification service (Kurtz, Eichhorn, Accomazzi, Grant, Henneken, Thompson, et al., 2003), the model matches the data very well, suggesting that the usage obsolescence function attributable to professional astronomers has been very stable over the past six years.

The obsolescence obtained from Google users is very different in shape. The first five years after publication have slightly elevated usage, after which usage is lower and essentially constant, the $H$ component in equation 1. The obsolescence curve from Google Scholar use is even more different: It rises as a function of age (Figure 1.7); we examine this in the next section.

This example, as well as work by Bollen and Van de Sompel (2008), makes clear that usage obsolescence, as well as essentially any other usage statistic, is critically dependent on the nature of the users. Whereas citation obsolescence is caused by the actions of scholarly authors, usage obsolescence has no such a priori set of users. The fact that the universe of users of scholarly articles can be much broader and different from the universe of scholarly authors presents both substantial challenges and substantial opportunities.

## Comparison with Citations

The relation between usage and citations is complex (e.g., Line & Sandison, 1975); it depends on both the nature of the document being used/cited and the nature of the individuals performing these actions. That different disciplines have different practices and mores concerning citation behavior has been known for quite some time (Burton & Kebler, 1960; Krause, Lindqvist, & Mele, 2007; Moed, Van Leeuwen, & Reedijk, 1998); Figure 1.6 shows that different sets of users can exhibit substantially different behaviors even when accessing the same documents. Indeed, comparing the obsolescence curves for the standard Google interface in Figure 1.6 with the Google Scholar interface in Figure 1.7 demonstrates that the specific behavior of the search engine can have a significant effect, as can the databases in which a paper is indexed (Hood & Wilson, 2005).

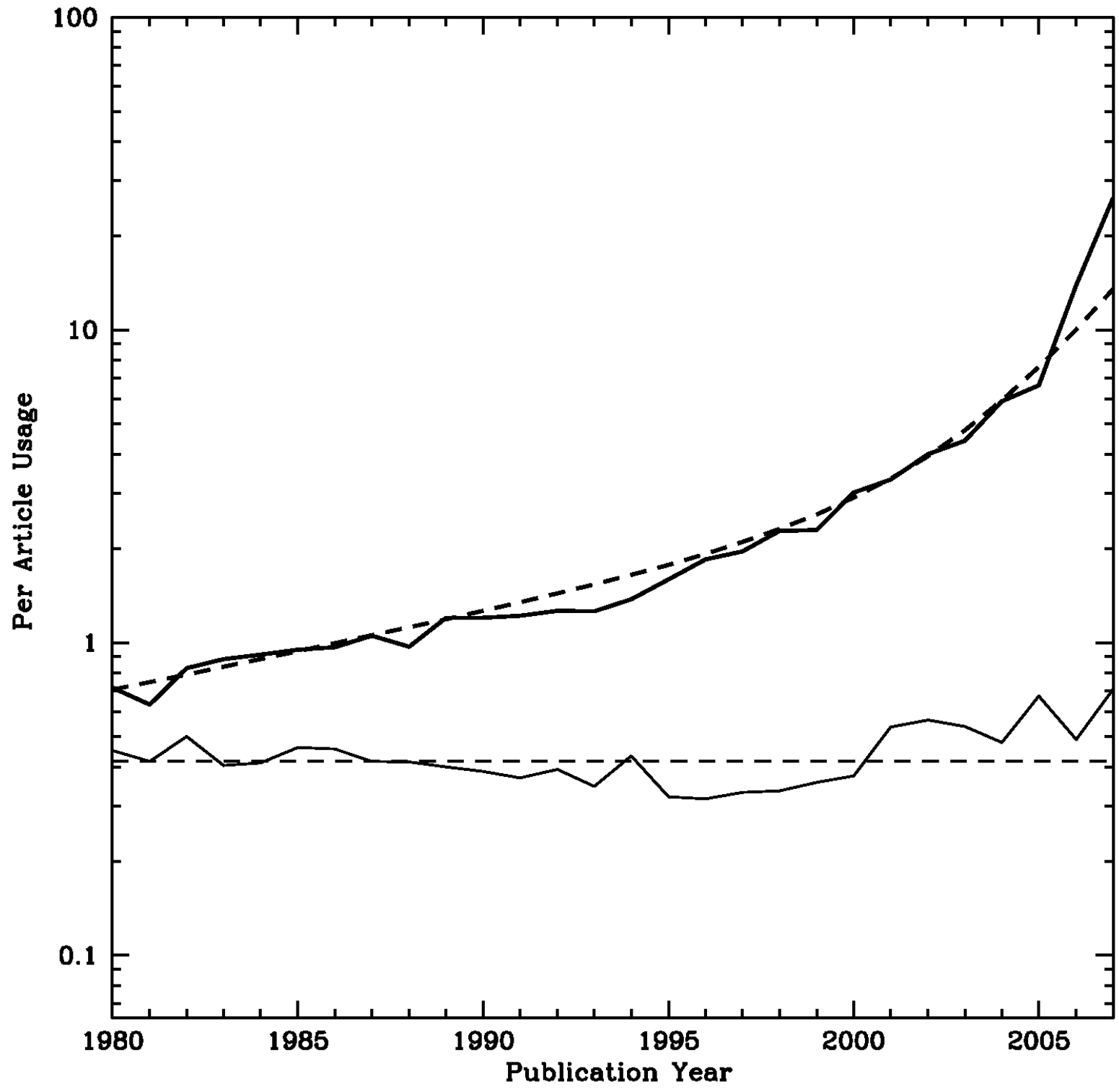

**Figure 1.6 Obsolescence of articles from the four main astronomy journals (*Astronomical Journal*, *Astronomy and Astrophysics*, *Astrophysical Journal*, and *Monthly Notices of the Royal Astronomical Society*) by frequent ADS users and by Google users in terms of actual use as a function of publication date. The top thick solid line represents use in October 2007 by individuals who used the ADS query engine and downloaded 10 or more articles during that month; typically these are professional astronomers. The lower thin solid curve represents the use by individuals who come directly from the Google search engine. The top dashed curve is exactly the *R* model of equation 1 and Figures 1.4 and 1.5; the bottom dotted line is a constant, independent of publication date.**

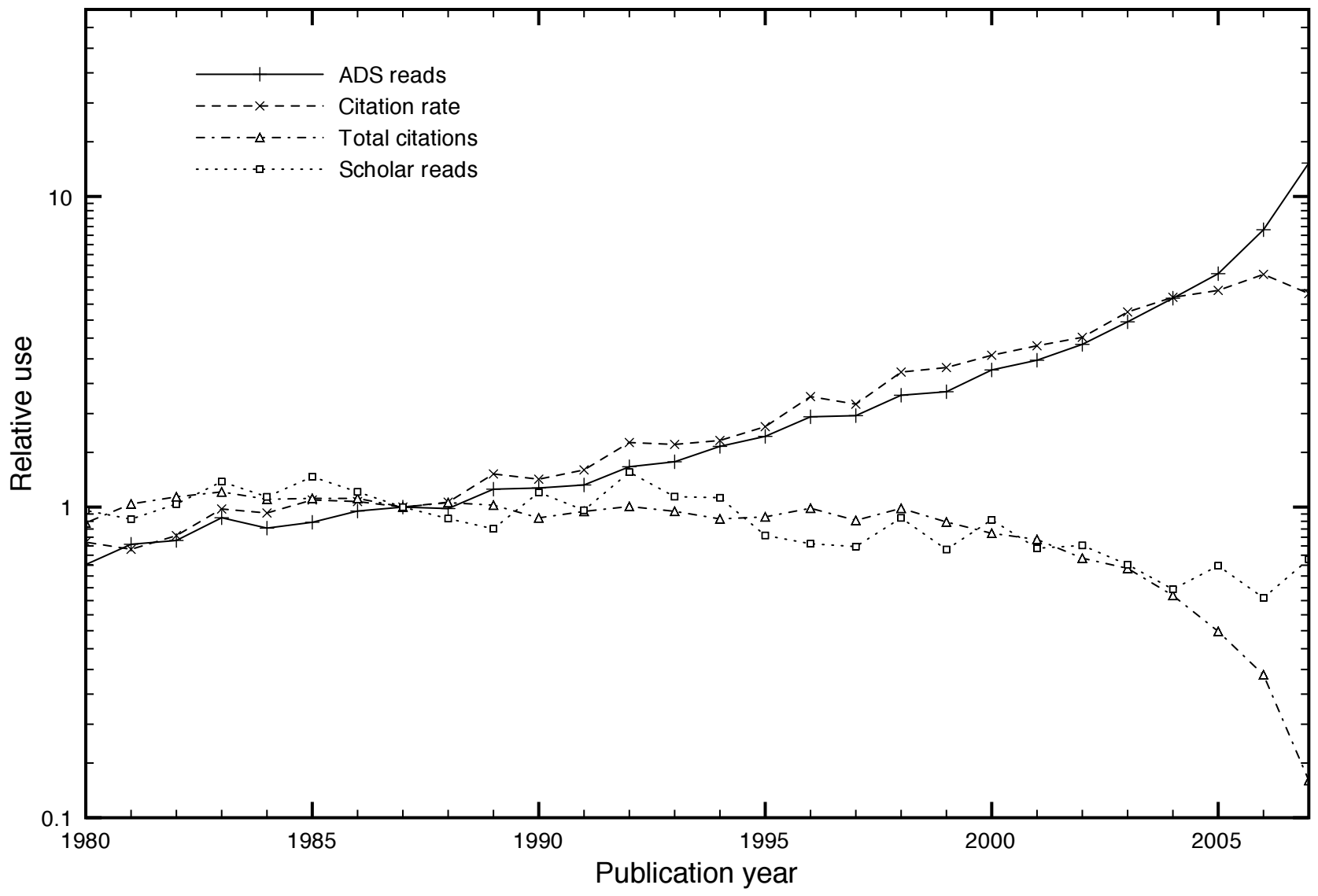

**Figure 1.7 Obsolescence of astronomy articles by frequent ADS users and by Google Scholar users. The top line (marked with X) shows use as a function of publication date for the main astronomy journals (*Astronomical Journal*, *Astronomy and Astrophysics*, *Astrophysical Journal*, and *Monthly Notices of the Royal Astronomical Society*) by heavy ADS users (mainly research astronomers). The thicker line (marked with +) is the citation rate to these articles. The line on the bottom (marked with boxes) shows the use of these articles by users who came to the ADS site via Google Scholar, and the line marked with triangles shows the total citations to these articles. All four functions are on a per article basis, and all four have been normalized to one for the publication year 1982 by dividing each function by its respective value for 1982; this normalization permits the shapes of the functions to be easily intercompared (Henneken et al., 2009).**

Of crucial importance to any comparison of usage with citations is the exact specification of the documents involved in the comparison. Garfield (1972) pointed out that newspapers are used a great deal but are almost never cited; thus it would make little sense to have a usage/citation study based on newspapers. Similarly, the *Bulletin of the American Astronomical Society* publishes mainly unrefereed conference abstracts; the *Astrophysical Journal* is the society's flagship journal, providing full length refereed papers. Papers in the *ApJ* are cited about 300 times more frequently than abstracts in the *BAAS*, but they are read only about six times as often. There is thus a factor of fifty difference in the usage to citation ratios for the two journals; clearly there are many measurements where it would not be appropriate to merge the two.

Kurtz, Eichhorn, Accomazzi, Grant, Demleitner, Murray, and colleagues (2005) pointed out that the series of papers by Trimble and colleagues: "Astrophysics in XXXX" (where XXXX is the immediate past year) each year is routinely the most used paper in astronomy, but is very rarely cited; for example, the most recent two, (Trimble, Aschwanden, & Hansen, 2006, 2007) have yet to be cited (as this is written). These papers are like newspapers in their usefulness. Clearly care must be taken, even for papers within a single journal, in not misconstruing effects.

An additional factor in determining the use of an article is that articles are read because they are cited. Kurtz, Eichhorn, Accomazzi, Grant, Demleitner, Murray, and colleagues (2005) showed the strong correlation between total citation and use for older papers, and Moed (2005) discussed the relationship for newer papers. Some search interfaces, such as Google Scholar, emphasize citation counts in their rankings. Figure 1.7, from Henneken, Kurtz, Accomazzi, Grant, Thompson, Bohlen, and colleagues (2009), shows how this affects the obsolescence curve. The top, solid line is the use of the main astronomy journals by heavy ADS users, the same curve as in Figure 1.6. The thick dotted line is the citation rate to these articles, measured during the three months following the measurement of use (November 2007–January 2008); the bottom thin solid line represents usage of the main astronomy journals by individuals who come to ADS via Google Scholar and the thin dotted line represents the total citations to these articles. The curves are normalized to the number of articles published and to the values for 1987, to emphasize the differences and similarities in shape.

We are thus justified in adding another, the fifth, independent function to the description of usage obsolescence $R_S$ to indicate the student's or learner's use:

$$R(t) = R_H + R_I + R_C + R_N + R_S \quad (2)$$

where $R_H$, $R_I$, $R_C$, and $R_N$ are as above and $R_S$ is proportional to the total citation count:

$$R_S = S_0 \int_{t_0}^{0} C dt$$

where $C$ is the citation rate and $S_0$ is a proportionality constant.

With the addition of the S mode we have a description of usage obsolescence that represents five different information seeking behaviors (and this is for only astrophysics journal articles). It may be assumed that the amplitudes and decay constants for these modes vary with a number of factors, including time, academic discipline, metadata composition, search engine performance, type of usage event, and mix of user types. Nicholas and colleagues (2005) list three types: practitioner, researcher, and undergraduate. We would add also the interested public. Eason, Richardson, and Yu (2000) identify eight user types.

As has been clear for some time (e.g., Gardner, 1990; Harnad, 1990) the advent of electronic research publication has had, and will continue to have, enormous influence on which articles are read, how they are read, and by whom. In the past, articles could be read only by researchers with access to scholarly libraries, more recently by those at subscribing scholarly institutions. Now, with changing cost and payment structures (such as Open Access [Drott, 2006]) an increasingly broad population of interested individuals can and does make use of the scholarly research literature.

With the accurate description of use being so complex, it is perhaps not surprising that the relation between use and citation has not been convincingly established. Parker (1982) first noticed that the use obsolescence function required the sum of two exponentials, similar to the citation obsolescence functions of Burton and Kebler (1960); matching parameterizations was impossible, because Parker needed to use a very broad selection of materials in order to achieve a significant result, whereas Burton and Kebler showed that the parameterization depended critically on the exact specification of research field.

Direct comparisons over the same set of input documents are rare; Line and Sandison (1975) were among the first to discuss the issue. Tsay (1998a) reviewed the pre-electronic results, essentially agreeing with the previous result of Broadus (1977, p. 319): "There do seem to be parallels between use of materials (not limited to journals) as indicated by citation patterns and as shown by requests to libraries." Stronger statements did not appear to be warranted. Scales (1976) correlated the top 50 journals according to use with their citation ranking and found a statistically significant correlation; however, the correlation of the top 50 journals according to citations with the ranked list of these journals by use was not significant. Brookes (1976) criticized Scales's work on methodological grounds; Bensman (2001) later confirmed Scales's findings; Meadows (2005) discusses this controversy. Pan (1978) and Rice (1979) found similar results. Stankus and Rice (1982) come to a key conclusion (quoting Tsay's [1998b, p. 32] paraphrase), stating that "a correlation will exist if the following conditions are met: (1) comparisons are made only among journals of fairly similar scope, purpose, and language; (2) with respect to the correlation between the citation data for a journal and the use of that journal, only if there is heavy journal use in that specialty or library."

Tsay's (1998a) results agree quite well with those of Stankus and Rice (1982): There is a significant correlation between the citations to medical journals and the use of these journals in a medical research library, but no significant correlation for the citations to other journals with use in a medical research library.

Past information comparing the actual shape of the obsolescence function for usage versus citations is very rare; again Tsay (1998b) summarizes and completes the pre-electronic results. Tsay found that use half-life was significantly shorter than citation half-life; Guitard in 1985 (discussed by Line, 1993) found use half-life longer than citation half-life, and Cooper and McGregor (1994) found use half-life shorter than citation half-life. King, McDonald, and Roderer (1981) found the data for various fields mostly not comparable.

Clearly usage and citations are not exactly correlated. Several studies have been done, and essentially each comes to a somewhat different conclusion. The difficulty in comparing usage information with citation histories is likely the underlying reason behind the controversy over the normative theory of citation, which would predict (MacRoberts & MacRoberts, 1987) that, in the aggregate, citations are proportional to use (e.g., Baldi, 1998; Cronin, 1984, 2001, 2005; Liu, 1993).

Toward the end of the twentieth century there was a rapid change in usage, as electronic media replaced paper. Kurtz and colleagues (2000) investigated use obsolescence in the ADS usage logs and found the four components of equation 1; they also compared use obsolescence with citation obsolescence, and found, for the period available, an exact match (Figure 1.8).

The data available to Kurtz and colleagues (2000), from mid 1998, were insufficient to examine any correlation between the very short term $R_N$ or the long term $R_H$ usage modes. By 2002 data became available to test the long term $R_H$ mode correlation (Kurtz, Eichhorn, Accomazzi, Grant, Demleitner, Murray, et al., 2005); no relation was found (Figure 1.9), leading to the equation relating use and citation obsolescence:

$$C(t) = c(R_C + R_I)(1 - e^{k_D}) \tag{3}$$

where $R_C$ and $R_I$ are as in equation 1, $c$ is a proportionality constant (the mean number of times an article is cited per use) and $k_D$ parameterizes the citation latency (e.g., Egghe & Rousseau, 2000), which has been getting substantially shorter in the electronic era (Brody, Harnad, & Carr, 2006).

The relation between the $R_N$ usage mode and citations is more subtle and clearly cannot yield to the direct comparison of mean behaviors done by Kurtz and colleagues. Moed (2005, p. 1088) found a correlation coefficient between article downloads during the first two months and total citations after 25 months of only 0.11; he suggests that "initial downloads and citations relate to distinct phases in the process of collection and processing relevant scientific information." Brody and colleagues (2006) also examined the correlation of early downloads versus citations for individual articles and found significant correlations, but they looked at a substantially longer time interval; essentially mixing the $R_N$ mode with the $R_C$ mode. Moed's (2005) result also showed larger correlations for longer time windows.

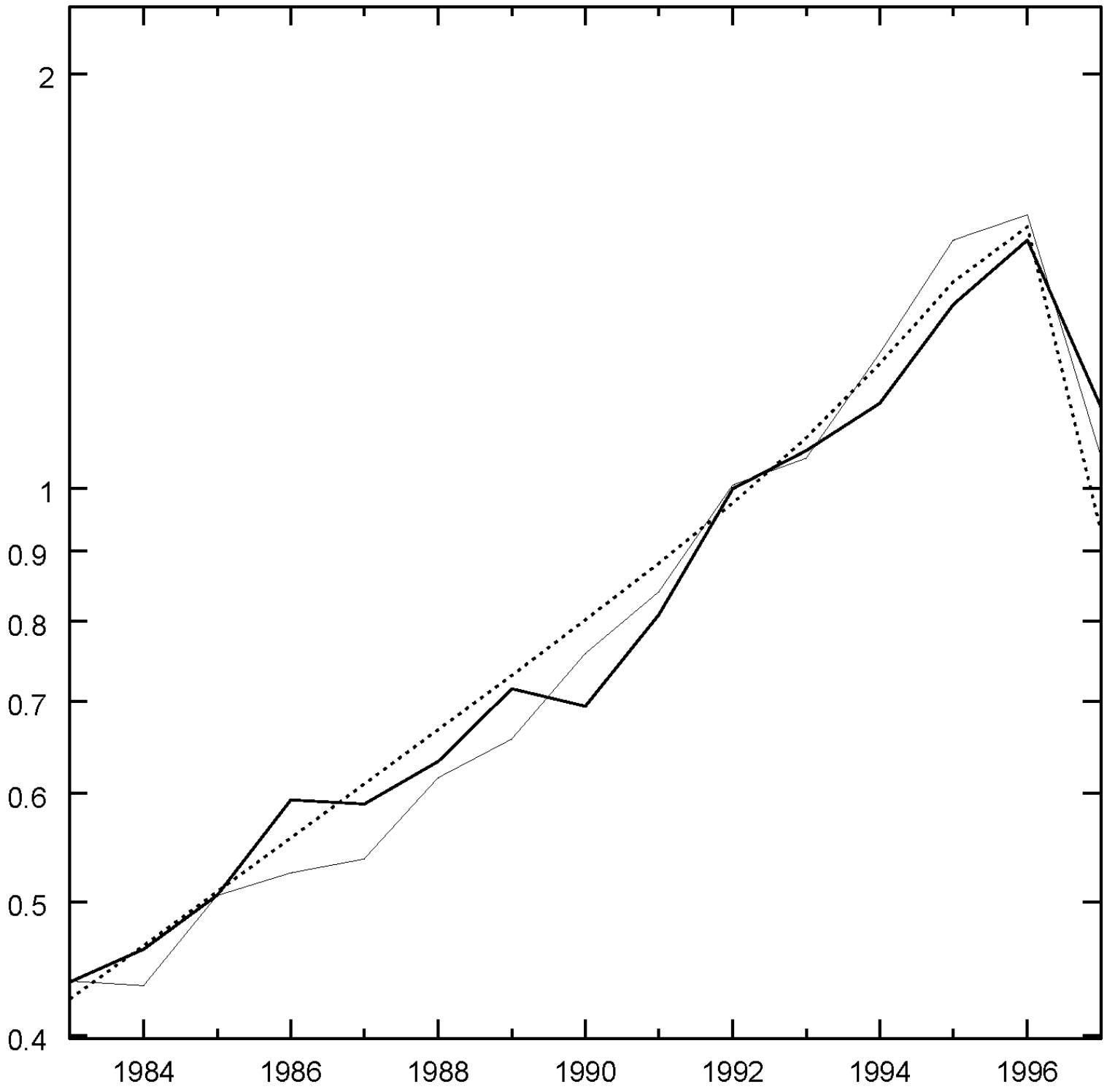

**Figure 1.8 Citation obsolescence of astronomy articles compared with actual use and the ($R_I$+$R_C$) model of equation 3. The thick solid line represents the actual citation rate to these articles for the first nine months of 1998 (shown on the y axis) as a function of publication date (shown on the x axis), on a per article basis. The thin solid line represents the actual use of these articles through the ADS interface during this same time period, times the exponential ramp up factor of equation 3. The dotted line is the model citation rate, *C* from equation 3 (Kurtz et al., 2000).**

Both Moed (2005) and Brody and colleagues (2006) examined the correlation between usage and citation rates to individual articles within the first few years after publication. Kurtz, Eichhorn, Accomazzi, Grant, Demleitner, Murray, and colleagues (2005) examined this correlation for two sets of data: All the usage of the entire ADS database in 2000 and the use in 2000 of *Astrophysical Journal* articles published from 1990–1997. For the larger sample they found that the citation rate was a good predictor of usage, but not the reverse (use did not predict citations). They attributed this to two main causes: Much of the use of the database is due to articles too new (the $R_N$ mode) to be cited, and to documents that are rarely cited. The *Astrophysical Journal* study showed a much stronger relationship between usage and citation. The time period covered is one where the $R_I$ mode dominates. There, use predicted citation just as well as citation predicted use. The relation in both directions was approximately log-normal, with a standard deviation of about a factor of two. Log-normal distributions have now been shown to be common in a number of similar social dynamic situations, such as multi-party voting (Castellano, Fortunato, & Loreto, 2009; Fortunato & Castellano, 2007).

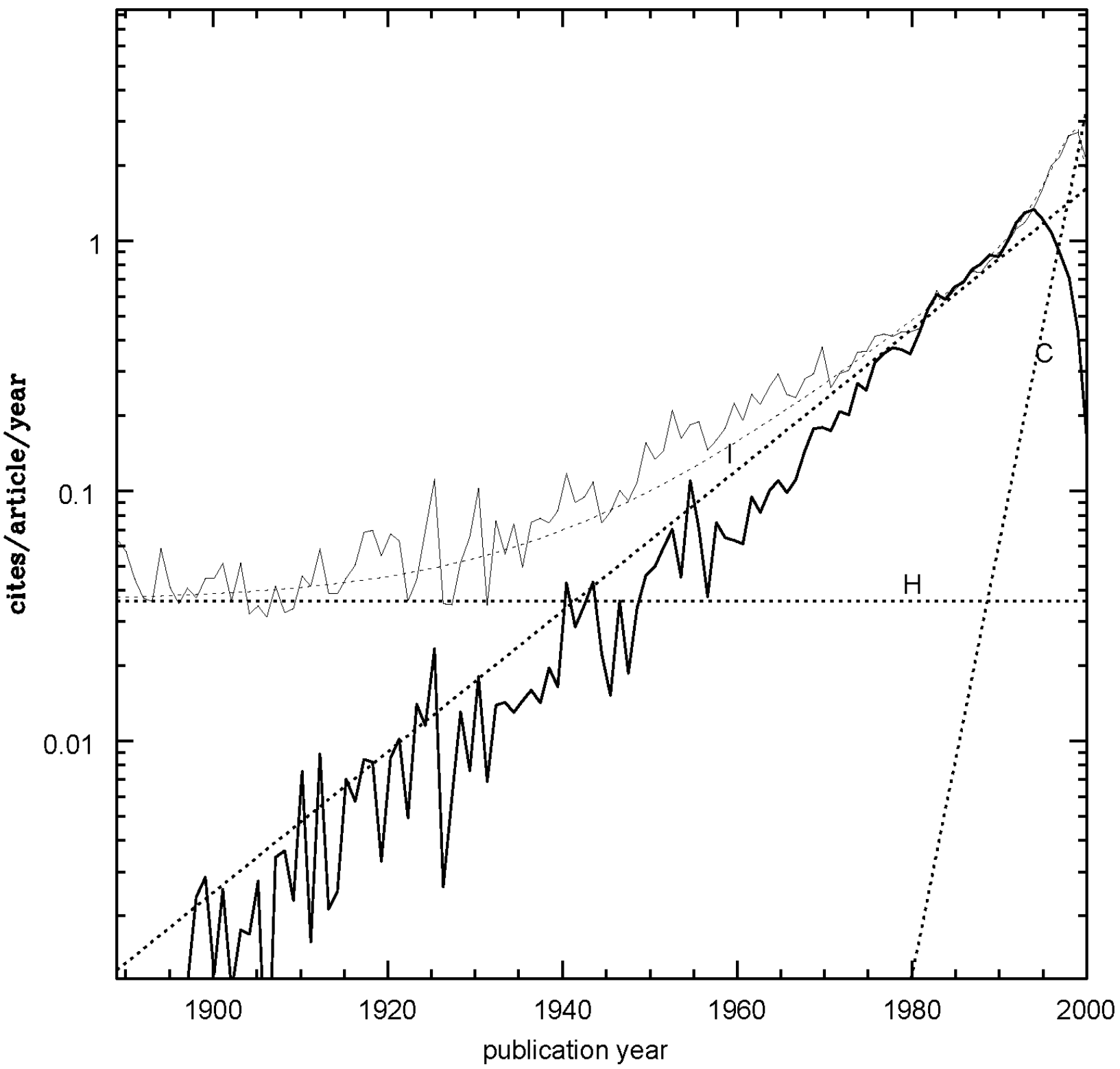

**Figure 1.9 Long term citation obsolescence of astronomy articles compared with actual use and the ($R_I$+$R_C$) model of equation 3. Notice that the long term constant, $R_H$ usage does not correspond to the citation behavior (Kurtz, Eichhorn, Accomazzi, Grant, Demleitner, Murray, et al., 2005).**

### Summary

Although usage obsolescence (usage frequency as a function of publication date) is perhaps the most straightforward possible measure of use next only to total use, its interpretation is clearly quite complex. In the half century since Burton and Kebler (1960) and Price (1965a) demonstrated the need for a two component model to describe citation obsolescence it has not been necessary to add any additional complexity to the basic model; usage obsolescence, in contrast, has (at least) five components (equation 2). Citation measures are mediated by the actions of the researchers who write scholarly articles; usage measures are not due solely to the actions of the authors. Although writers are often users, users can often be non-writers. That the number of usage events vastly exceeds the number of citation events provides the basic information that will allow the complexity of the usage information to be understood; herein lies the future of the types of measurements and analyses discussed in this chapter.

## *Some Usage-Based Statistical Measures*

Like citation or publication counts, usage rate information can be used to measure the effectiveness and productivity of a number of entities, such as authors, journals, academic departments/universities, and countries. Unlike citation or publication count information, the use event records contain information not only about the entity (e.g., article) being used, but also about the user. This permits entirely new types of measures. Some general issues concerning these are discussed by Bertot, McClure, Moen, and Rubin (1997); Jansen (2006); Mayr (2006); Peters (2002); and Rowlands and Nicholas (2007).

By the nature of their time dependencies, usage data are sensitive to the recent publication record of an individual or organization but citation data are sensitive to the full integrated history.

Both traditional and usage bibliometrics can aggregate measures based in the properties of an article yielding, for example, number of citations or downloads to articles where country X is the address of the author; usage bibliometrics can also aggregate user based measures, such as downloads from users in country X. These techniques can, obviously, be merged.

Before describing some examples of how this is being done currently, we should make explicit some caveats about the use of these measures. As has been discussed, the populations that use articles can be very different from the populations that cite articles; this adds substantial complications to the understanding of the results of a usage based analysis, complications that should be understood before these measures are used in employment or funding decisions.

Also complicating the situation is the possibility of cheating, manipulating the data to influence a decision. As an example it would take only a couple of extra downloads per article per week for a tenure candidate to move from average to outstanding. Because use is mostly anonymous this may be impossible to detect.

## Individual Article

Per article use is a direct measure of the popularity of an article; co-use is often used as a measure of similar content. It has become a common feature in many digital libraries to show co-use statistics for an article, for example, "People who read this article also read …" The ADS system has, since 1996, allowed the use of "second order operators" (Kurtz, 1992; Kurtz, Eichhorn, Accomazzi, Grant, & Murray, 1996, 2002; Kurtz, Eichhorn, Accomazzi, Grant, Demleitner, & Murray, 2005) that operate on the attributes of ensembles of articles. A common use is to begin with a collection of articles on a single subject and return a list of articles that are currently most popular with persons who "also-read" articles in the original list; this tends to yield the "hottest" papers on a subject.

As with any collaborative filter (e.g., Goldberg, Nichols, Oki, & Terry, 1992), one must decide exactly whose opinions should count. The ADS filters out infrequent users, thus following Nicholas and colleagues (2005) it might be better to include only those who download the full text.

## Authors

The main usage modes of researchers, $R_C$+$R_I$, are well correlated with citation counts for individual articles, as has been discussed. The two measurements, although both related to the usefulness of an article, have very different properties. Usage rates decrease monotonically with time following publication, even as citation counts increase monotonically. Usage rates are a measure of the current use; citation counts are a measure of all past use.

By taking the combined citation counts and usage rates for an aggregation of papers by a single author one obtains a two dimensional measure of that author's productivity or usefulness, which, in addition to the author's age, gives substantially more information than citation counts alone when evaluating performance.

Figure 1.10, from Kurtz, Eichhorn, Accomazzi, Grant, Demleitner, Murray, and colleagues (2005), shows how this works. The points represent individual research astronomers, and the positions of the points show the total citations to each author's papers and the amount of use those papers received (in the ADS system) between January 1999 and May 2000 (all measures normalized by number of authors in each paper).

The solid line in Figure 1.10 represents the locus of a very productive researcher's life in this diagram; the locus for less productive individuals would move to the left and would not reach as high. In addition to an individual's productivity the model includes retirement and the growth in the number of papers published with time.

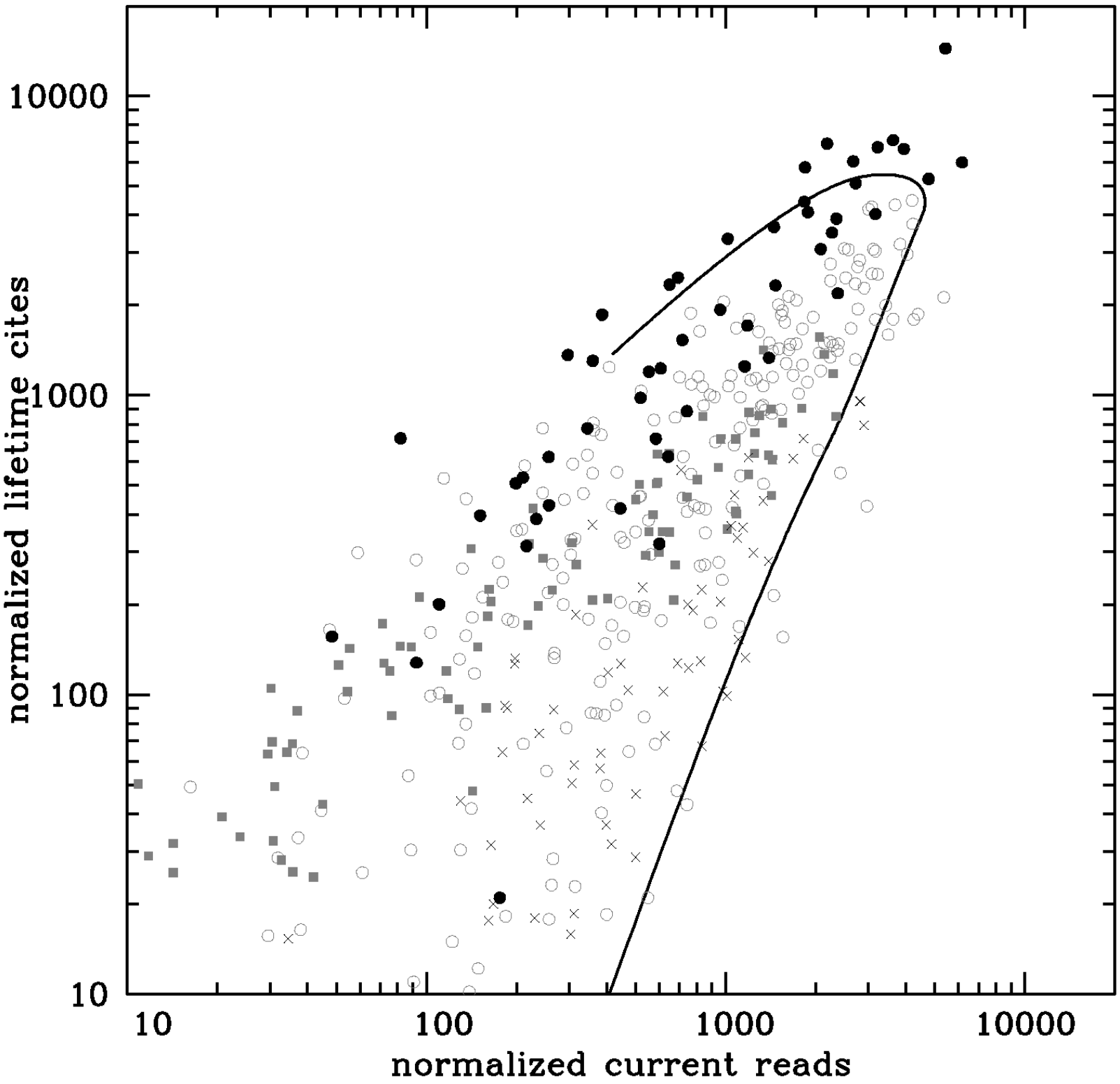

**Figure 1.10** Usage rate vs. total citations for individual astronomers; the solid line is a model for the most productive scientists at different ages (Kurtz, Eichhorn, Accomazzi, Grant, Demleitner, Murray, et al. [2005] fully describe the plot and the model).

Figure 1.11 shows how, in addition to the length of time past the Ph.D. the "read-cite" diagram can be used to help evaluate an individual's performance. The solid dots represent the total usage rate and total citation counts for persons who received their Ph.D.s in astronomy from U.S. universities in 1980 and who published at least one item in the astronomy literature after 1990. The three sets of vectors represent models for these individuals for three different productivities (the lowest is ten times less productive than the highest), and the two vectors (for each productivity level) represent two life histories, where one individual stops publishing after ten years and the other continues publishing the entire twenty years.

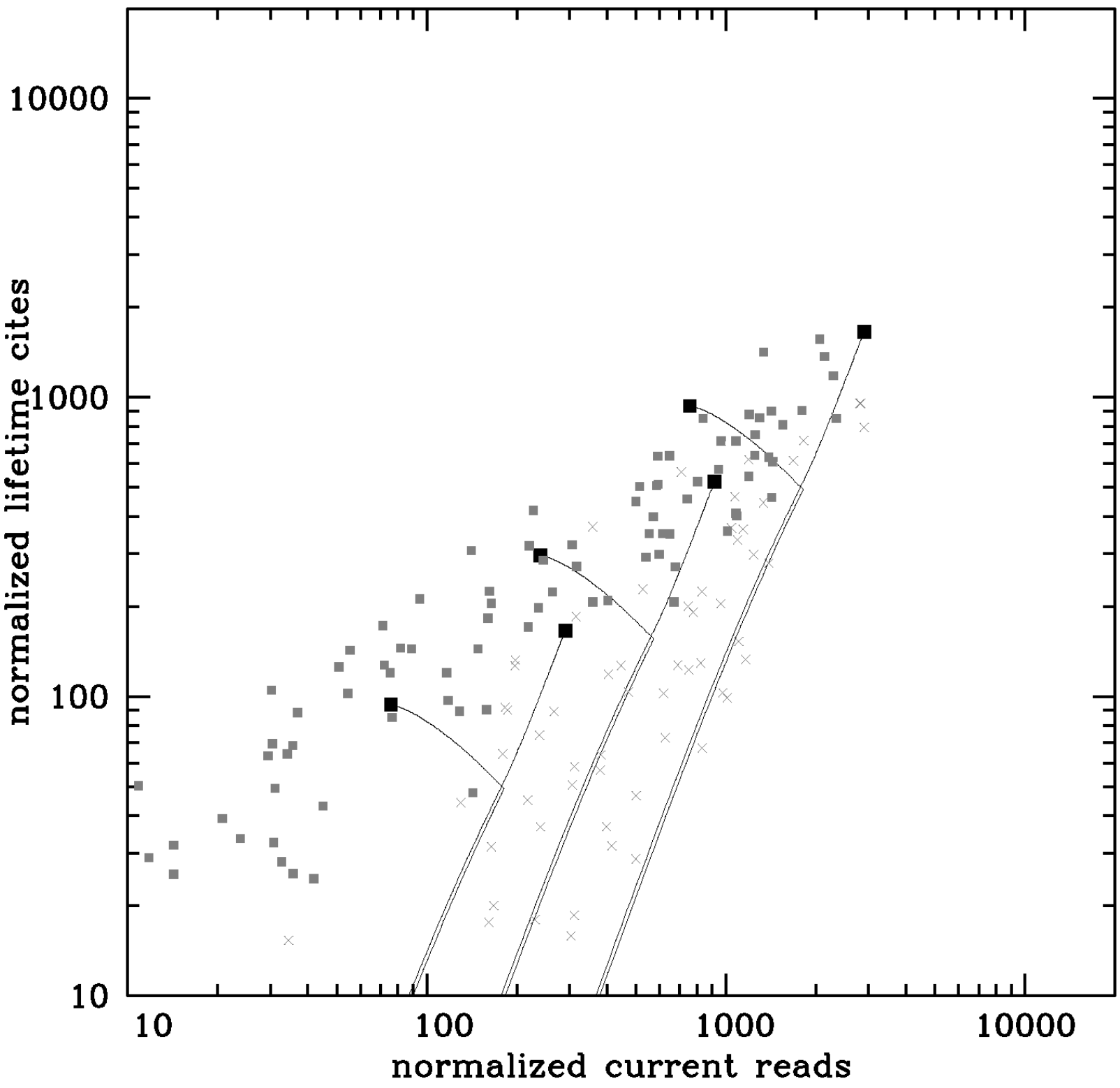

**Figure 1.11** **The solid lines represent different productivity levels and work histories for astronomers in the usage rate total citation diagram (Kurtz, Eichhorn, Accomazzi, Grant, Demleitner, Murray, et al. [2005] fully describe the plot and the models).**

## Journals

Librarians have made purchase decisions on the basis of local usage data for a very long time (e.g., Gross & Gross, 1927; Walter & Darling, 1996). This has continued into the electronic era; for example, see articles by Darmoni, Roussel, and Benichou (2002); Davis (2002); and several papers in Fowler's (2007) book.

With the advent of electronic journals, usage studies have gone beyond the local use of journals. With all these studies the application and acceptance of the results depend crucially on the relevance of the sample. As with the rest of usage bibliometrics, studies involving journal use are in their infancy; there is no set of measures that is even remotely as complete or accepted as the journal citation impact factor (Garfield, 1972).

Kurtz, Eichhorn, Accomazzi, Grant, and Murray (1997) and Kurtz and colleagues (2000) studied differences in the shape of the usage obsolescence curves for several core astrophysics journals; they found indications of changes in editorial policy, differences in the relative currency of journals, and usage rate differences as a function of the user's country of origin.

Kaplan and Nelson (2000) looked into changes in the local citation rate as a function of the usage rate for documents that were included in an electronic library versus print-only documents; they found no effect. In a similar, much larger study McDonald (2007) was able to show significant increases in the local citation rate, correlated with the online availability and use of journals. In contrast, Kurtz, Eichhorn, Accomazzi, Grant, Demleitner, Henneken, and colleagues (2005) found no effect on the citation rate for older astronomy research articles after they had been digitized and posted, free, on the Internet.

Bollen, Luce, Vemulapalli, and Xu (2003) calculated centrality values for journals with a network where journals form the nodes and co-use forms the links. (Two journals would be linked to each other in the network if the same person downloaded an article from each of them.) In addition to finding obvious differences between the Thomson Reuters' citation impact factor for journals and the usage patterns for these journals in a physics research lab (Los Alamos National Laboratory was the source of the usage logs) they found they could show changes in research focus over short time scales (three years).

Davis and Price (2006) examined how usage statistics are affected by electronic journal design, making cross comparisons of different journals and publishers difficult. Blecic, Fiscella, and Wiberley (2007) similarly showed that the evolution of both interface and search methodologies is differentially changing the detailed meaning of usage statistics, again making cross comparisons difficult.

Bollen and Van de Sompel (2006a, 2008) have proposed a "Usage Impact Factor" similar to the Citation Impact Factor (Garfield, 1972). They have examined some of the sample biases that affect these measures; Bollen, Van de Sompel, and Rodriguez (2008) have analyzed and classified several different measures of journal quality or impact, both citation based and use based (see Figure 1.17).

## Departments

Universities and university departments may be viewed as collections of individuals; persons who are both creators and users of scholarly information. It is possible to measure both the use of articles written by members of a university faculty and the use of articles by members of a university faculty. The latter presents obvious problems of privacy.

Very little work has been published exploring these data. Kurtz, Eichhorn, Accomazzi, Grant, Demleitner, and Murray (2005) compared the number of heavy users of the astrophysics literature at a sample of major astronomical research centers with the number of members of the American Astronomical Society at those centers to support their claim that nearly every astronomy researcher is a heavy ADS user; similar usage information is included in usage reports from publishers to libraries.

Kurtz, Eichhorn, Accomazzi, Grant, Demleitner, Murray, and colleagues (2005) examined the use of articles written by members of several prominent astronomy research faculties, both combining and contrasting usage measures with citation measures. Figure 1.12, for example, shows the number of authors in each of the highly ranked faculties who are in the top third (among all tenured faculty in the top departments) in terms of total citation (y-axis) and usage rate (using ADS log data) for articles published in the most recent ten years (x-axis, what Kurtz et al. call Read10). This is basically a prestige versus activity diagram.

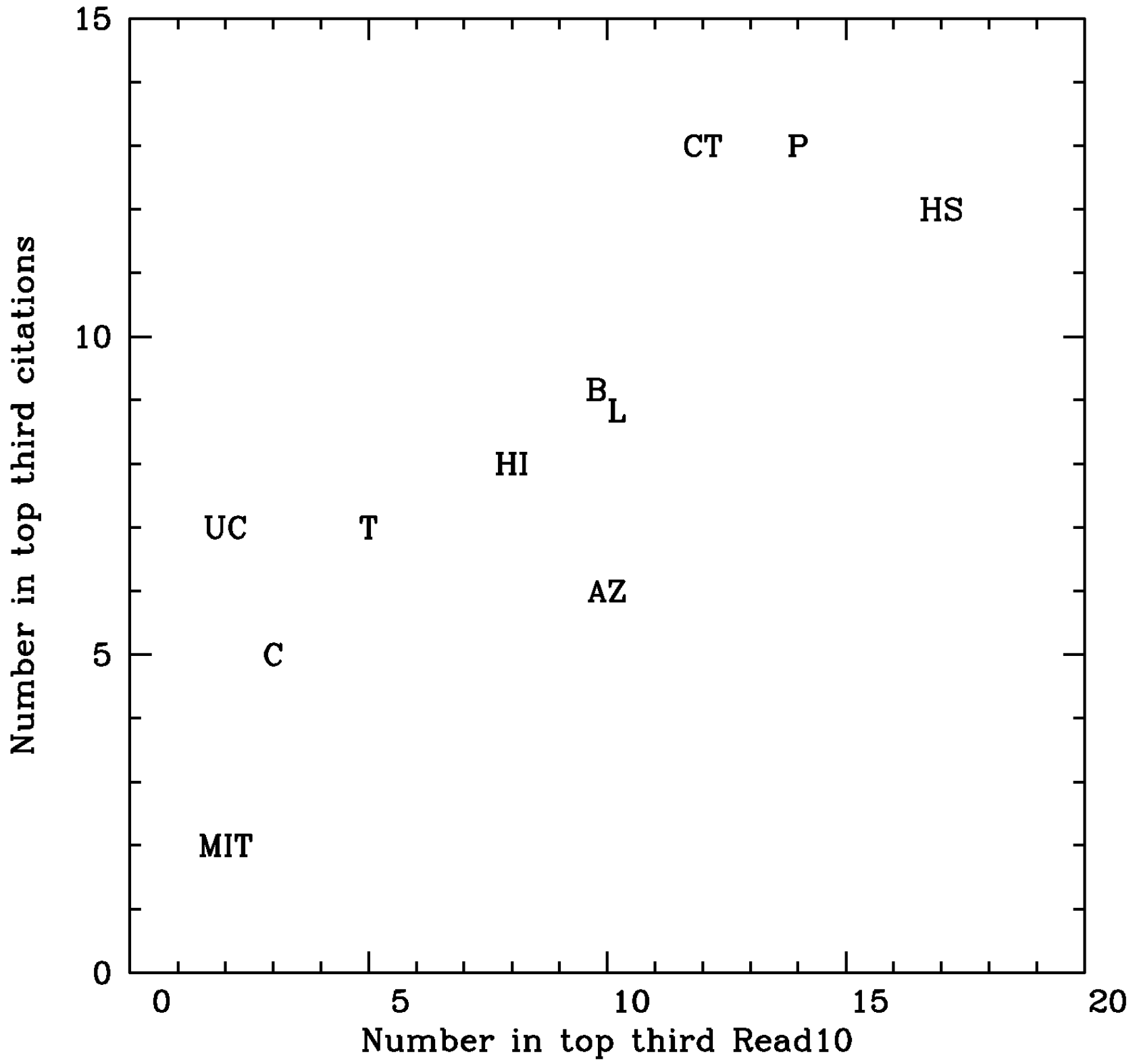

**Figure 1.12 A prestige vs. activity diagram for the top U.S. astronomy faculties (Kurtz, Eichhorn, Accomazzi, Grant, Demleitner, Murray, et al. [2005] fully describe the plot and the symbols).**

If the practical problems associated with gathering and interpreting usage event data can be solved, it may be expected that measures such as those shown in Figure 1.12 will become standard means for assessing the academic status of groups.

## Countries

Comparing the scholarly output of countries has been an important branch of bibliometrics for decades (e.g., Leydesdorff, 2008; May, 1997; Price, 1965b); the number of papers attributed to authors from a country, and citations to these papers are routinely tabulated in the Thomson Reuters Essential Science Indicators (sciencewatch.com).

To our knowledge, no systematic study to date has compared countries according to the usage of articles originating in them. A small number of studies has compared countries according to the number of scholarly usage events originating in them (Eichhorn et al., 2000; Henneken et al., 2009; Kurtz, Eichhorn, Accomazzi, Grant, Demleitner, & Murray, 2005).

Figure 1.13, from Kurtz, Eichhorn, Accomazzi, Grant, Demleitner, and Murray (2005), shows the relationship between per capita use of the astrophysics literature (from the ADS logs) and per capita Gross Domestic Product (GDP) for several countries. The straight line represents a quadratic relationship; per capita use is proportional to per capita GDP squared. King (2004) found a similar result, using citations.

Figure 1.14, from Henneken and colleagues (2009), shows the relative growth in use of the astrophysics literature by selected countries and regions as a function of per capita GDP; the factor of 30 total growth in global use over the ten years covered by the plot is divided out to show the relative growth. Interestingly, after very rapid comparative growth China is now growing at only the world average; India, on the other hand, continues to grow faster than the rest of the world (which is dominated by Europe and North America).

One property of usage information is that it is available in near real time; one can know the results of a usage data analysis, in some cases, essentially as the use happens. This is also true for data concerning the number of publications, but not for citation information. That the use of scholarly literature approximately scales with the number of scholars times the per capita GDP (Kurtz, Eichhorn, Accomazzi, Grant, Demleitner, & Murray, 2005) allows these effects to be removed, making it possible to analyze the effects of changes in science policy independent of changes due to increasing GDP.

The traditional bibliometric measures available immediately scale linearly with GDP. Taking astrophysics as an example: The number of astronomers is a constant times the GDP of a country (Kurtz, Eichhorn, Accomazzi, Grant, Demleitner, & Murray, 2005); the number of papers from a country is a linear function of the GDP of that country (Krause et al., 2007) for the closely related field of high energy physics; and the number of papers is a linear function of the number

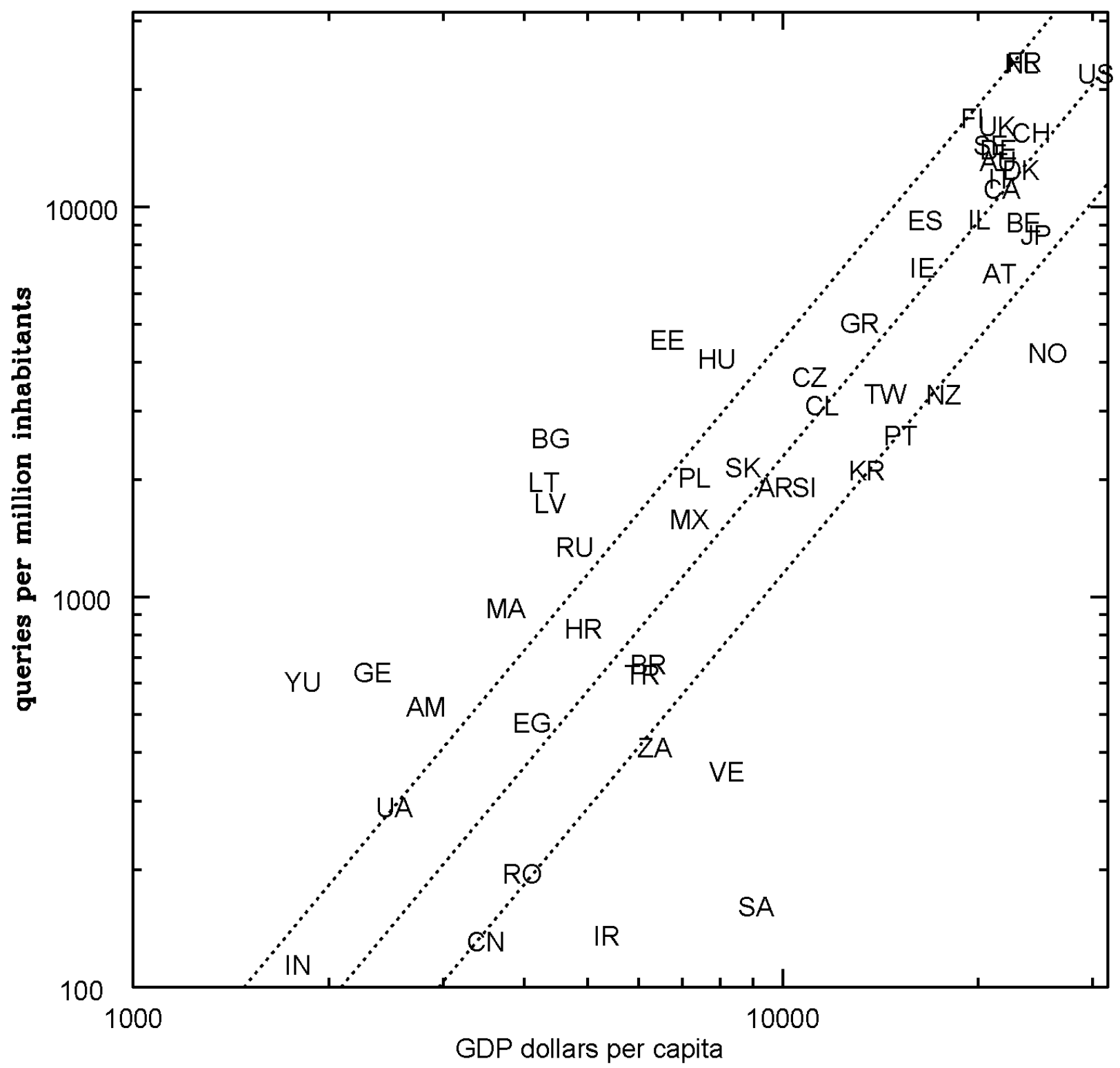

**Figure 1.13 The per capita use of the astrophysics literature vs. per capita GDP for several countries; the straight line represents a quadratic relationship (Kurtz, Eichhorn, Accomazzi, Grant, Demleitner, & Murray, 2005).**

of astronomers according to Abt (2007), who also showed the linearity for other disciplines.

That the use of scholarly articles by scholars in a country (Kurtz, Eichhorn, Accomazzi, Grant, Demleitner, & Murray, 2005) as well as citations to articles written by scholars in a country (King, 2004) both scale as a quadratic power law in GDP per capita suggests measurable differences that could, perhaps, be dubbed quality. If Kurtz, Eichhorn, Accomazzi, Grant, Demleitner, and Murray (2005), based on analyzing the outliers in Figure 1.13, are correct, GDP is a proxy for the infrastructure that supports research: Roads, universities, telecommunications systems, and so on. This leads to considering usage statistics as measures of the cumulative advantage of countries (Price, 1976) or, as Kurtz, Eichhorn, Accomazzi, Grant, Demleitner, and Murray (2005) point out, the Matthew effect for countries (Bonitz, 1997; Merton, 1968).

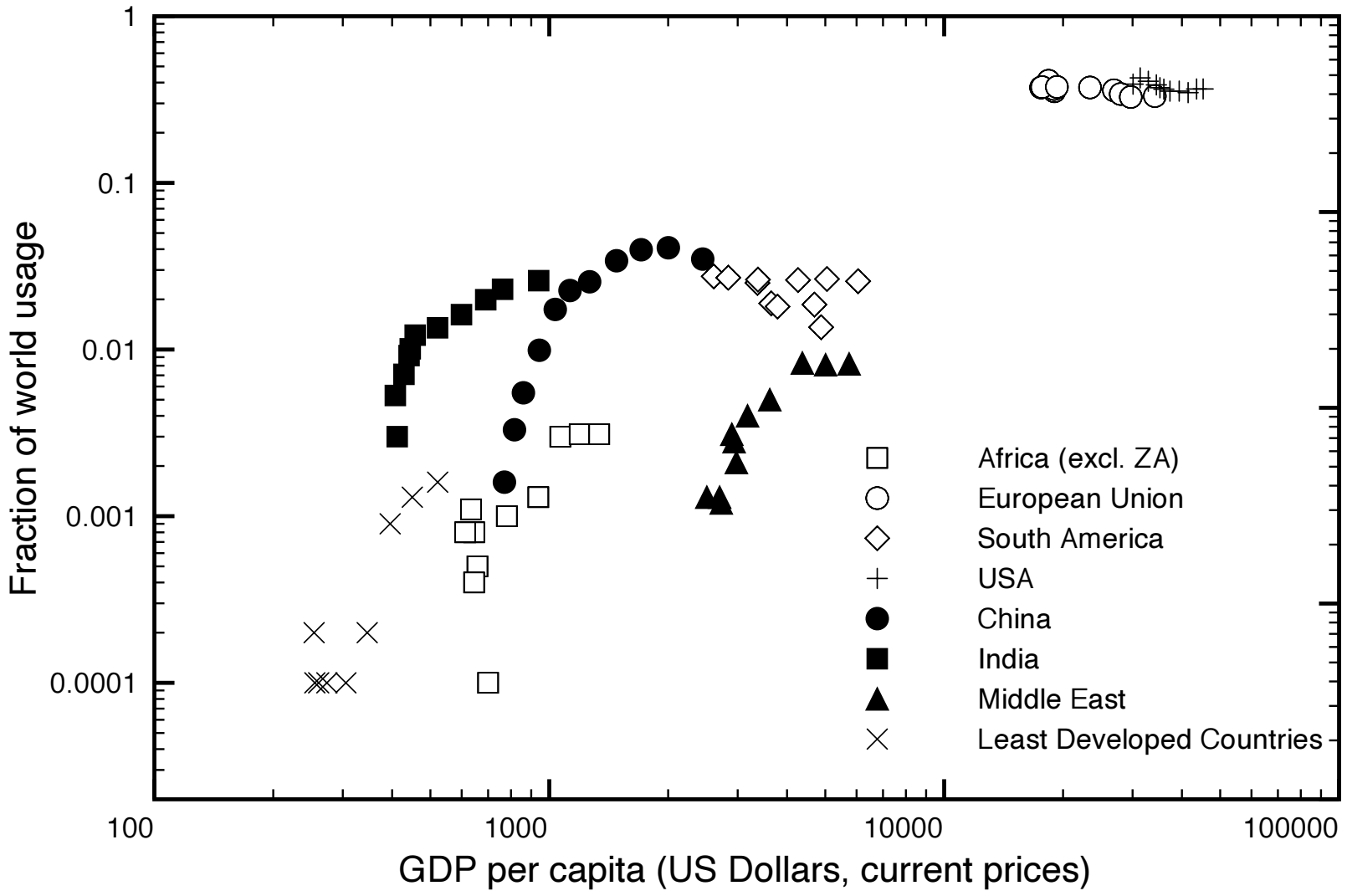

**Figure 1.14 The relative growth of use of the astrophysics literature by selected countries; each point is a single year's data (Henneken et al., 2009).**

## *Social Network Measures*

The results mentioned previously indicate that usage-based measures can be used to asses the impact of authors, articles, and journals. Usage rates furthermore predict future citation rates and show other interesting relationships to citation data. However, research also indicates that usage-based measures are sensitive to distribution parameters and sampling characteristics. In addition, questions can be raised with respect to their semantics. Do normalized usage and citation rates, such as the Usage Impact Factor and Citation Impact Factor, effectively express impact? Or do they rather express notoriety or popularity?

This situation is similar to that encountered in the study of social phenomena. To assess the status of individuals one could simply record the rates of their social interactions, for example, the number of e-mails, telephone calls, or endorsements received. However, such measures would be fraught with distribution and sampling biases and provide only very partial indications of individual status. The most common approach, therefore, is to consider social status a relational phenomenon that emerges from the structure of an individual's relationships. Social network analysis has generated numerous measures of individual status of this nature (Wasserman & Faust, 1994), for example, eigenvector centrality and betweenness centrality.

## Social Network Measures for Citation Data

Social network analysis has been applied sporadically to scholarly assessment. Various social network measures can be calculated on the basis of citation networks to assess journal and article impact. This approach has received considerable interest in recent years, but has not yet been expanded to usage bibliometrics. We provide an overview of its basic principles and measures, in particular in the context of the most commonly applied social network measures of node status, and subsequently discuss their applications to usage bibliometrics.

## Degree Centrality

Assume we collate all citations that pass from one journal's articles to another journal's articles in terms of the $n \times n$ matrix $A$. $n$ thus corresponds to the number of journals for which we have citation data. The entries of this matrix, $a_{i,j}$ , are 1 if any number of citations point from journal $v_i$ to $v_j$ and 0 if no citations point from journal $v_i$ to $v_j$. A similar formulation can be developed for instances where

$$a_{i,j} \in \mathbb{N}^+$$

corresponds to the number of citations that point from journal $v_i$ to $v_j$.

Matrix $A$ now represents a citation network among $n$ journals. From this citation network we can define a set of social network measures of journal status.

The degree centrality measure follows the rationale of measures such as the Impact Factor, namely that the impact or status of a journal is defined by its rate of endorsement. However, because matrix $A$ is asymmetric, that is,

$$a_{i,j} \neq a_{j,i}$$

endorsements can be calculated as "incoming" or "outgoing" of a particular journal $v_i$. We thus define the in-degree centrality of journal $v_i$ according to matrix $A$ as

$$C_{\text{in}}(v_i, A) = \frac{\sum_j a_{j,i}}{\sum_i \sum_j a_{i,j}} \quad (4)$$

This definition can be modified to normalize

$$C_{\text{in}}(v_i, A)$$

by the maximum in-degree over matrix $A$ or other normalization factors. Note that this definition holds for either definition of $A$ where its entries are

$$a_{i,j} \in 0, 1 \quad \text{or} \quad a_{i,j} \in \mathbb{N}$$

Likewise, we can define the out-degree of a journal $v_i$ according to matrix $A$ as

$$C_{\text{out}}(v_i, A) = \frac{\sum_j a_{i,j}}{\sum_i \sum_j a_{i,j}} \tag{5}$$

The definition of the Impact Factor is similar to that of in-degree centrality; with the exception that

$$\sum_i \sum_j a_{i,j}$$

is replaced by the number of articles published in journal $v_i$, hence an average per-article citation rate.

### Eigenvector Centrality, PageRank, and Random Walk Measures

In-degree and out-degree centrality, like all other rate-based measures of journal impact, suffer from one particular deficiency. They count the number of citations regardless of where they originate. As such they serve as an indicator of journal "popularity" but not necessarily of its impact, "influence," or "prestige." An example of this situation could be a journal whose articles receive numerous citations from lower-ranked journals (popular) in contrast with a journal whose articles receive few citations but from highly prestigious journals (prestigious).

The seminal work by Pinski and Narin (1976) first introduced the notion that journals should be ranked according to not only the number of citations they receive, but also whether these citations originate from influential journals. In particular, citations that originate from highly ranked journals should have a greater weight in determining the status of the cited journal. However, this introduces a circular reasoning; how, then, do we define the influence of the citing journals, and so on? The answer lies in the mathematically guaranteed convergence of an iterative recalculation of journal influence values.

Assume we define an influence vector $\vec{p}$ such that each entry $p_i$ corresponds to the influence of journal $v_i$. We can now define the influence value of each journal $v_i$ as the sum of the influence values of the journals by which it is cited, namely

$$p_i \simeq \lambda \sum_{v_j \in C(v_i)} \frac{p_j}{N(v_j)} \tag{6}$$

where $C(v_i)$ is the set of journals that cite journal $v_i$, $N(v_j)$ the number of journals cited by journal $v_j$, and $\lambda$ a linear scaling parameter. The $p_j$ values are normalized by $N(v_j)$ so that $v_i$ receives a portion of $p_j$ according to the total number of journals that $v_j$ cites; as $v_j$ cites more journals, each receives a lesser portion of its influence. This expression roughly corresponds to the definition of PageRank as originally proposed by Brin and Page (1998), namely

$$PR(v_i) = \frac{(1-\lambda)}{N} + \lambda \sum_j PR(v_j) \times \frac{1}{N(v_j)} \tag{7}$$

where $N$ represents the total number of journals. This definition simulates a random walk over the citation network wherein

$$\frac{(1-\lambda)}{N}$$

also referred to as "teleportation" factor, adds a probability of random teleportation.

We can reformulate equation 6 on the basis of matrix $A$ as:

$$p_i \simeq \lambda \sum_{j=1}^{n} \frac{p_j a_{j,i}}{N(a_{j,>0})} \tag{8}$$

where $N(a_{j,>0})$ corresponds to the number of non-zero entries of the row vector $a_j$.

We can reformulate the above expression in terms of the matrix-vector multiplication

$$\vec{p}_t = \lambda A \vec{p}_{t-1} \tag{9}$$

so that $\vec{p}_t$ gives the influence values of all journals at iteration $t$. Using the power iteration method $\vec{p}_t$ will converge to the primary eigenvector of $A$ such that

$$A\vec{p} = \lambda \vec{p} \tag{10}$$

redefining the issue as an eigenvector problem. Several different methods can be used to determine the primary eigenvector of $A$. However, the efficiency of the PageRank algorithm when calculated for very large, highly sparse graphs such as the Web's hyperlink network has caused it to become the preferred means to approximate the rankings of journals according to their citation eigenvector centrality.

The Eigenfactor project (www.eigenfactor.org) uses a principle similar to PageRank to produce journal rankings. Bollen, Rodriguez, and Van de Sompel (2006) propose using a weighted version of Google's PageRank algorithm to rank journals according to their prestige. Results indicate a significant, domain-dependent deviation of journal rankings according to Thomson Reuters' Impact Factor. The scatterplot in Figure 1.15 demonstrates how some journals, in particular review journals in medicine, receive high rankings according to their Impact Factors but receive low PageRank scores. This pattern is an indication of a journal receiving many citations, but from relatively unprestigious sources. Vice versa, journals characterized by low Impact Factor values but high PageRank values can be said to receive fewer citations but from more prestigious sources.

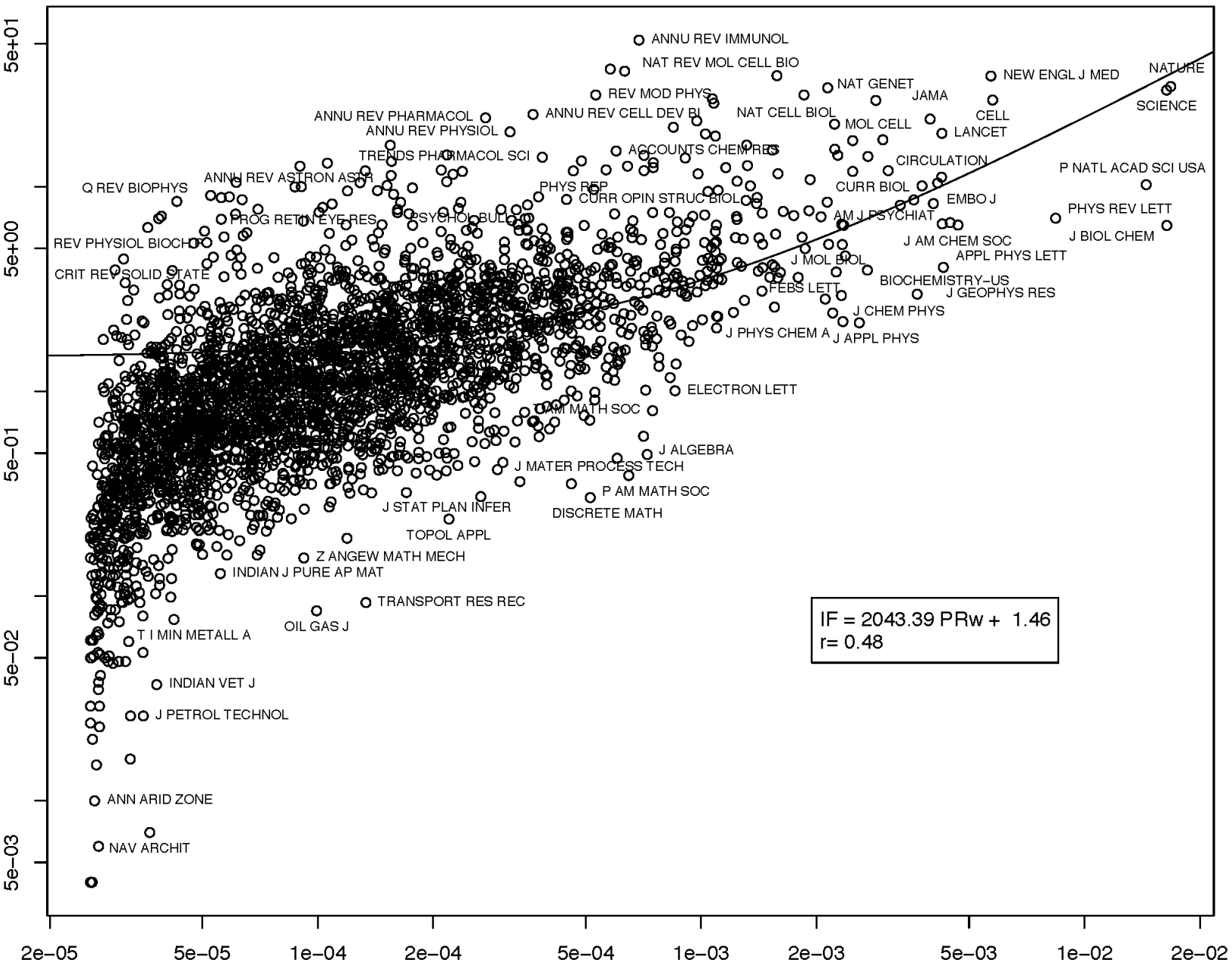

**Figure 1.15 Comparison of journal impact factors and PageRank values calculated from 2003 *Journal Citation Reports* citation network (Bollen, Rodriguez, & Van de Sompel, 2006)**

## Shortest Path Measures

Shortest path social network measures, including betweenness and closeness centrality, rely on the notion of a citation geodesic, that is, the shortest path connecting a pair of journals in a citation network. Bollen, Van de Sompel, Smith, and Luce (2005) apply betweenness and closeness centrality to a citation network to produce journal rankings, further explored by Bollen and colleagues (2005, 2008) and Leydesdorff (2007), who demonstrated that citation betweenness can serve as an indicator of journal interdisciplinarity.

We can formalize closeness and betweenness centrality as follows. We denote the geodesic between journals $v_i$ and $v_j$ in the network represented by the adjacency matrix $A$ as the ordered set

$$g_{i,j,A} = (v_i, \cdots, v_j) \tag{11}$$

The set of all geodesics in $A$ we denote

$$G(A) = \{\forall(v_i, v_j) | g_{i,j,A}\} \tag{12}$$

The length of a geodesic in $A$ is given by:

$$L(g_{i,j,A}) = ||g(v_i, \cdots, v_j)|| - 1 \tag{13}$$

In case the entries of matrix $A$ correspond to normalized citation counts, that is,

$$a_{i,j} \in \mathbb{R}^+ \quad \text{and} \quad \textstyle\sum_i a_{i,j} = 1$$

we can define a geodesic weight that corresponds to the joint product of the normalized citation counts as

$$P(g_{i,j,A}) = \prod_{k=1}^{k=L(g_{i,j,A})} a_{v_k,v_{k+1}} \tag{14}$$

where $a_{v_k,v_{k+1}}$ represents the entry of $A$ that corresponds to the journal pair

$$v_k \in L(g_{i,j,A}) \text{ and } v_{k+1} \in L(g_{i,j,A}).$$

The journal closeness centrality of journal $v_i$, denoted $C(v_i,A)$, can now be defined as the mean length of all geodesics originating in $v_i$:

$$C(v_i, A) = \frac{\sum_{g \in G(A)} L(g_{i,j,A})}{||G(A)||} \quad (15)$$

In other words, closeness centrality expresses how far removed a journal is from all other journals in the citation network and thus how central and important it is to the network. Journal betweenness centrality of $v_i$, denoted $B(v_i, A)$, is then defined as a function of the number of times a particular journal sits on the geodesic of any pair of journals, that is,

$$B(v_i, A) \simeq \lambda ||\{v_m, v_n : v_i \in g_{m,n,A}\}|| \quad (16)$$

where $\lambda$ is a normalization factor. Betweenness centrality expresses how crucial the journal is in establishing the connections among other journals in the network and thus its interdisciplinary "power" position in network structure. Removing journals with high betweenness centrality from a citation network would break up and interrupt the paths that connect many pairs of journals.

## Social Network Measures for Usage Bibliometrics

Usage data consist of sequences of temporally ordered usage events that do not inherently contain journal to journal or article to article networks as citation data do. Therefore it is not possible to calculate network-based measures of impact from usage data without further processing. However, networks of resource relations can be extracted from usage data using methods that determine resource relationships by examining the degree to which pairs of resources occur within the same user sessions. Such methods are related to association rule learning in data mining (Aggarwal & Yu, 1998; Mobasher, Dai, Luo, & Nakagawa, 2001). They rest on the assumption that if users frequently issue requests for the same pairs of resources within the same session, referred to as co-usage, these resources may be statistically related. The degree of relationship is commensurate with the relative frequency by which the pair of resources is *co-used*.

Bollen and colleagues (2005) first proposed the extraction of resource networks from usage data and the subsequent calculation of network-based measures of journal or document status. This method can be summarized as shown in Figure 1.16.

Assume we have a log data set $R$ of $k$ requests $r \in R = \{r_1, r_2 ..., r_k\}$. Each request $r = \{d, t, s, q\}$ where $d$ is a document identifier, $t$ a request date-time, $s$ a session identifier, and $q$ a request type. We now assign each request $r$ within a session $s$ a rank order

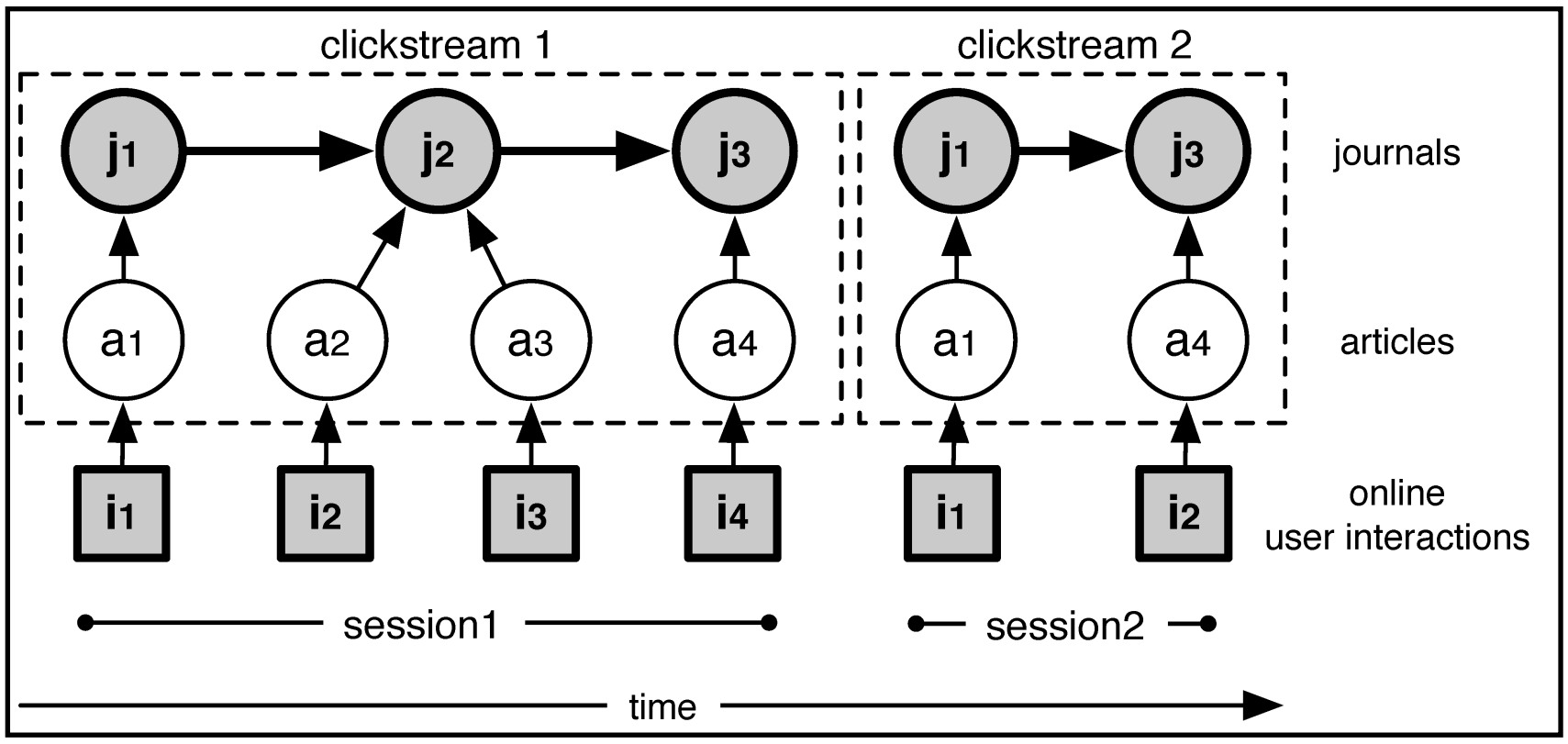

**Figure 1.16 Extraction of journal networks from usage data**

$$O(r, s) \in \mathbb{N}^+$$

according to its date-time stamp, so that we can order the requests in a session according to the sequence in which they took place. We extract a set of request pairs

$$T = \{(r_i, r_j) \in R \times R : O(r_i, s) = O(r_i, s) - 1\}.$$

We now define the function

$$F : F(d_i, d_j) \rightarrow \mathbb{N}^+$$

which returns the frequency by which each document pair $(d_i, d_j)$ occurs in $R$, so that $d_i \in r_i$ and $d_j \in r_j$ for the request pairs $(r_i, r_j) \in T$. $F(r_i, )$ returns the frequency of all pairs in which $r_i$ is the antecedent. This frequency can be normalized to determine the transition probability

$$p(r_i, r_j) = \frac{F(r_i, r_j)}{F(r_i, )} \quad \textbf{(17)}$$

$p(r_i,r_j) \in [0,1]$ and expresses the probability that users in their clickstreams will move on to journal $r_i$. Note that, because the temporal sequence of requests is taken into account, the established document relationships are directional, that is, $(r_i,r_j) \neq (r_j,r_i)$ unlike other association learning approaches that disregard temporal sequence and rely on a bidirectional definition of *co-occurrence*. The resulting transition probabilities define a matrix $P$ whose entries $p_{i,j} \in [0,1]$ represent the transition probability between documents $d_i$ and $d_j$ according to the set of request sequences in $R$. The measures discussed in the section on social network measures for citation data are thus applicable to matrix $P$ and ranking of documents can be determined according to their relationships as indicated from usage.

Bollen and colleagues (2005, 2008) discuss a journal ranking on the basis of journal usage networks created from Los Alamos National Laboratory (LANL) link resolver logs and the MESUR reference data set. The rankings indicate the ability of a variety of social network measures to express various features of scholarly impact. Deviations between the resulting rankings and the Impact Factor point to how social network measures calculated from usage networks indicate various measures of scholarly impact. The MESUR project reports Spearman Rank Order correlations between the various social network metrics of impact and the Impact Factor that range from 0.28 to 0.80. These indicate that social network metrics can exhibit various degrees of congruence with existing citation-based measures of impact.

Bollen and colleagues (2008, 2009b) visualize the difference between rankings produced by a set of 43 measures of journal impact, including the Impact Factor, by a principal component analysis (PCA) of the set of measure correlations. The first two principal components cover nearly 85 percent of all variation. The resulting PCA mapping of measures (shown in Figure 1.17) reveals the two main clusters of measures: Those based on usage data and those based on citation data. However, the second component separates measures according to the facet of impact they represent, namely the degree to which they represent prestige or popularity. The Impact Factor is positioned among degree centrality and closeness centrality metrics that are indicative of a journal's number of relations to other journals and thus its general degree of endorsements or its popularity. Betweenness centrality and PageRank cluster strongly as well, in two usage- and citation-derived clusters. However, citation- and usage-based betweenness centrality and PageRank measures correlate more strongly to each other than either does to the Impact Factor. The latter is placed at a position removed from the main clusters of prestige-based measures.

## *Open Access*

A recent, interesting bibliometric controversy concerns the relationship between usage and citations. Are articles that have been posted on the Internet for all to freely read more read/cited than articles that can only be read by subscribers? There is an extensive literature on this issue; Hitchcock's online bibliography (opcit.eprints.org/oacitation-biblio.html) currently has more than 200 items.

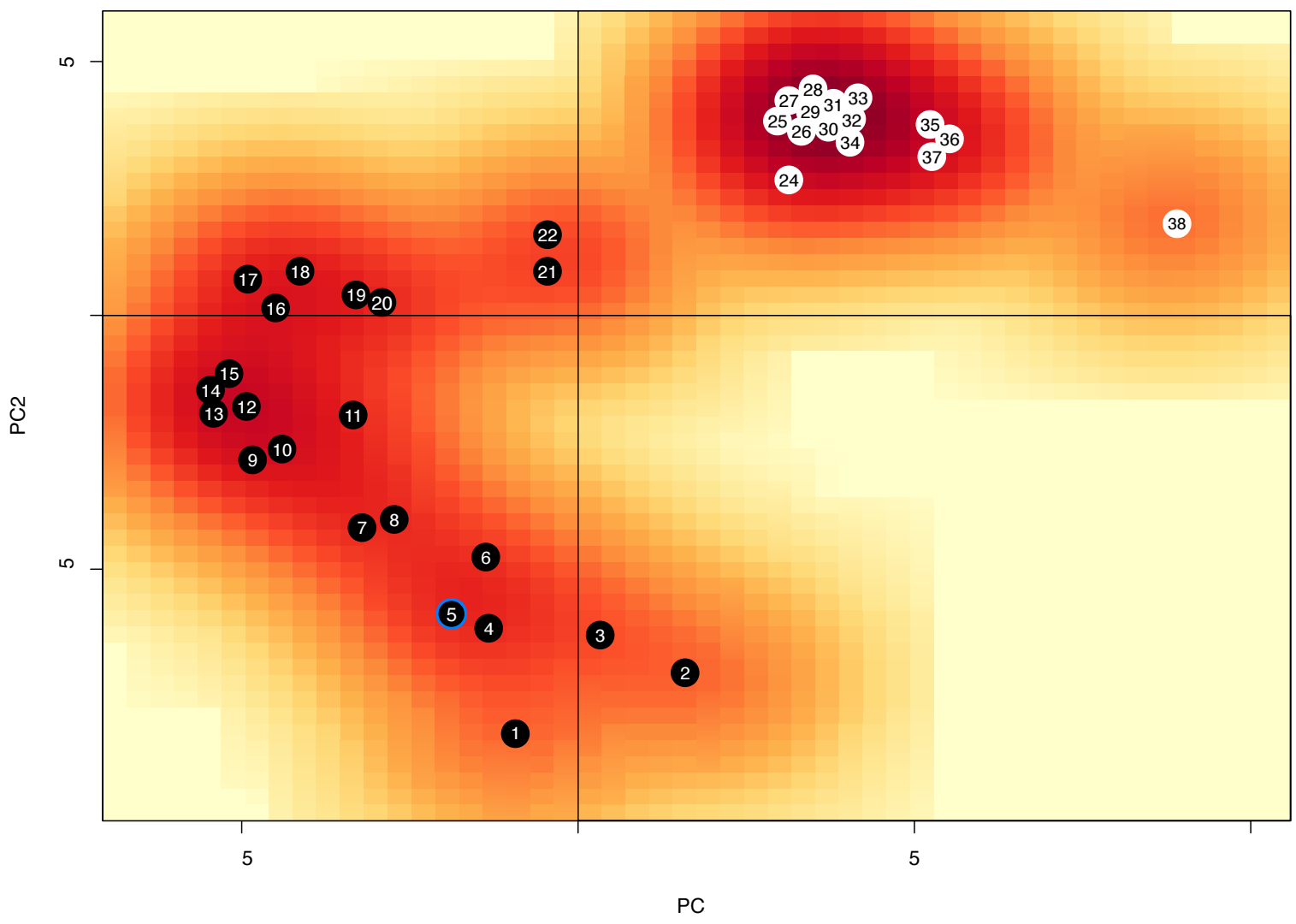

**Figure 1.17 Principal component analysis of the correlations between 39 impact measures calculated and retrieved by the MESUR project (Bollen et al., 2009b)**

Lawrence (2001) first showed that articles in computer science that had been posted on the Internet were cited at about twice the rate of similar articles that were not posted. Drott (2006) discussed some of the issues surrounding this observation. Harnad and Brody (2004) showed that the effect also existed, at about the same level, 2 to 1, in many of the subfields of physics covered by the arXiv e-print server (arxiv.org) (Ginsparg, 1994, 2001; Ginsparg et al., 2004). Combined with the strong correlation between usage and citation shown in Figure 1.8 and equation 1 from Kurtz and colleagues (2000; Kurtz, Eichhorn, Accomazzi, Grant, Demleitner, Murray, et al., 2005) and Kurtz (2004) showing that the ratio of full-text views to abstract views among astronomers is a strong function of ease of access to the full text, this led to the widely held assumption that the effect was causal: Because it is easier to access freely available articles than articles behind financial subscription barriers, more people read them and this *causes* them to be cited more frequently. This has been used as an argument in favor of open access (Eysenbach, 2006; Ginsparg, 2007) and is sometimes referred to as the "Open Access Advantage" (Harnad, 2005).

Kurtz, Eichhorn, Accomazzi, Grant, Demleitner, Henneken, and colleagues (2005) examined the question of causality more closely; their two main findings substantially undermined the causality postulate. First, they showed that older astrophysics articles that were posted on the Internet were not more frequently cited than they were before they were posted, in spite of substantially increased use; they did find evidence that very recent articles were more frequently cited following the founding of the arXiv e-print server. Second, they looked at astrophysics articles that were/were not posted on arXiv. They looked at three postulates explaining why arXiv posted articles were cited at roughly twice the rate of the articles not posted: The Open Access (OA) postulate, essentially the causal assumption of the previous paragraph; The Early Access (EA) postulate, articles are posted to arXiv several months (the median is four) before they appear in the journals, and this clearly gives them a head start in citation counts; and The Selection Bias (SB) postulate, more citable articles are more likely posted to the Internet than less citable articles, one possible reason (among many) for this being that authors preferentially post their better papers.

Kurtz, Eichhorn, Accomazzi, Grant, Demleitner, Henneken, and colleagues (2005) found that the distribution of highly cited papers was strongly inconsistent with the combined OA+EA postulates; this, combined with their historical analysis that showed no evidence for the OA postulate (but did show strong evidence for the EA postulate) led them to conclude that the increase in citations was due to a combination of EA+SB, with very small or no contribution from OA: That, in fact, the open access advantage did not exist.

Moed (2007) examined papers in condensed matter physics that were/were not posted on arXiv. He also found no evidence for the OA postulate and suggested that most of the effect was due to EA. Eysenbach (2006) showed that the effect existed in the "author choice" articles in the *Proceedings of the National Academy of Sciences*, where there could be no EA contribution. Henneken and colleagues (2006) showed that the use of the journal version of arXiv submitted articles vs. articles not submitted to arXiv in astrophysics closely parallels the citations and that amplitudes of the differences cannot be explained by EA alone. Davis and Fromerth (2007), looking at mathematics articles in arXiv, found no support for the OA or EA postulates, but found support for a quality bias (SB).

Craig, Plume, McVeigh, Pringle, and Amin (2007) reviewed the whole controversy, coming to the conclusion that the OA postulate was not necessary to explain the data and suggesting that a true randomized study be made. Kurtz and Henneken (2007) used the historical accident that the *Astrophysical Journal* switched from open access to closed access on January 1, 1998, to show that the arXiv/non-arXiv citation differential in astrophysics is associated only with arXiv submission, not with OA per se.

Subsequently, Davis, Lewenstein, Booth, and Connolly (2008), in collaboration with the journals of the American Physiological Society, conducted a true randomized statistical trial. Articles from APS journals were randomly assigned to be open access, or not, and their citation and usage were tracked. This experimental design totally eliminated any confusion from EA or SB factors. They found a significant increase in the full-text downloads for the open access articles but no difference in the citation rates between the open access and closed access articles, rejecting the OA postulate with high confidence. Note that both Harnad (2008) and Eysenbach (2008) have criticized this paper on methodological grounds.

Most recently, Norris, Oppenheim, and Rowland (2008, p. 1970) reviewed the controversy and found "the reasons why there is a citation advantage for OA articles still has not been satisfactorily explained"; Davis (2009), using methods similar to Eysenbach (2006), found that the effect could be explained by systematic differences between the OA/non-OA articles.

This change, from regarding the open access citation differential as causal to regarding it as a result of various biases, has not gone unchallenged. Stevan Harnad and his collaborators have provided the great bulk of the opposing view, primarily via online postings to the American Scientist Open Access Forum (amsci-forum.amsci.org/archives/American-Scientist-Open-Access-Forum.html) (Harnad is the moderator), the Sigmetrics listserver of the American Society for Information Science and Technology Special Interest Group on Metrics (listserv.utk.edu/archives/sigmetrics.html) (moderator: Eugene Garfield), and Harnad's own blog (openaccess.eprints.org).

These postings have provided a dialogue among the principals, often running for weeks. Representative postings, emphasizing the causal view, are by Harnad (2005; 2006a; 2006b; 2007a; 2007b; 2007c; 2007d). Perhaps the most complete exposition of their argument is in the paper by Hajjem and Harnad (2005).

Other important papers from this group include Hajjem, Harnad, and Gingras (2005) and Brody, Carr, Gingras, Hajjem, Harnad, and Swan (2007). Most recently Gargouri (2008), from an analysis of four mandated archives, found evidence for the causal hypothesis.

Although interesting, the importance of this issue is fleeting, as Ginsparg (2007, p. 16) points out:

> Studies have shown a correlation between openly accessible materials and citation impact, though a direct causal link is more difficult to establish, and other mechanisms accounting for the effect are easily imagined. It is worthwhile to note, however, that even if some articles currently receive more citations by virtue of being open access, it doesn't follow that the benefit would continue to accrue through widespread expansion of open access publication. Indeed, once the bulk of publication is moved to open access, then whatever relative boost might be enjoyed by early adopters would long since have disappeared, with relative numbers of citations once again determined by the usual independent mechanisms. Citation impact per se is consequently not a serious argument for encouraging more authors to adopt open access

> publication. A different potential impact and benefit to the general public, on the other hand, is the greater ease with which science journalists and bloggers can write about and link to open access articles.

In addition, as discussed in the section "Some usage-based statistical measures," if measures of scholarly impact can be extracted from usage data, the issue is not whether the impact of openly accessible materials increases because increased usage leads to increased citations, but that increased usage (and derived measures) itself is a sufficient indicator of increased impact.

## Mapping of Science from Usage Data

Other than ranking journals, authors, departments, and countries, usage data can play an important role in another crucial aspect of bibliometrics, namely, the mapping and charting of science. Maps of science highlight the main distinctions and structural trends in science by visualizing the connections among articles, journals, and scholarly domains. These connections are most frequently derived from citation data, for example, co-citation data and journal citation similarities.

### *Mapping Science from Citation Data*

Although even earlier attempts exist, Garfield (1970) provides an example of the usefulness of visualizing citation networks by invalidating the common misconception that Gregor Mendel's paper on genetics was ignored by that community. It is shown that Mendel's paper was cited repeatedly and was in fact not ignored. The flow of citations surrounding Mendel's paper is used to assess the flow of ideas and influences from one publication to another.

Small (1999) provides an excellent historical overview of the literature in this domain and shows discipline mappings resulting from the analysis of the citation relations between 36,720 documents. More recent and more large-scale efforts include Boyack, Wylie, and Davidson's (2002) use of the VxInsight (www.cs.sandia.gov/~dkjohns/JIIS/Vx_Intro.html) package to create landscapes of related papers in microsystems technology and the physical sciences. Temporal trends in this domain are revealed by highlighting shifts and peaks in the resulting landscapes. Boyack, Klavans, and Börner (2005) maps the relations among 7,000 journals on the basis of journal similarities calculated from Science Citation Index (SCI) and Social Sciences Citation Index (SSCI) citation data. Five citation-based similarity measures are calculated and journals are placed in a map of science by VxOrd, a force-directed algorithm to optimize journal positions on the basis of their similarity. The resulting maps of science outline the major structure of science activity in terms of 212 clusters of related journals.

Leydesdorff (1994) produces maps of scholarly domains by positioning journals by means of multi-dimensional scaling on the basis of 1993 SCI data. The citation environments of specific journals, for example, the *Journal of Chemical Physics*, are mapped in terms of groupings resulting from a factor analysis. The results indicate that the relations between the specific journals do not correspond to "the administrative division of the natural sciences into disciplines like physics and chemistry" (p. 65). The resulting visualization highlights important distinctions in the influences of the publications of a particular journal (or domain).

Moya-Anegón, Vargas-Quesada, Chinchilla-Rodrìguez, Corera-Álvarez, Munoz-Fernández, and Herrero-Solana (2007) discuss a "scientogram" of world science that was generated from co-citation data calculated for 219 Journal Citation Report (JCR) categories. The maps visualize the relations among various scholarly disciplines as provided by Thomson Reuters' classification codes.

Chen (2006) does not so much produce maps of science as concept maps that delineate the important distinctions in scientific domains. These data can be visualized in terms of decision trees that outline which terms are most important in determining the citation context of items in the Sloan Digital Sky Survey (SDSS) or concept maps generated by CiteSpace.

Rosvall and Bergstrom (2008) generate maps of science that group more than 6,000 journals into scientific domains by means of a random walk simulation on the citation graph; frequent paths can be increasingly compressed to ever more succinct descriptions of network structure. The resulting citation maps of science highlight the connections among various scholarly domains in terms of their citation flow. Similar maps are used in the eigenfactor.org online service.

Saka and Igami (2007) discuss research maps generated from co-citation patterns of highly cited papers extracted from SCI data. Hot research areas are identified by means of average annual growth rates of citations over a six year period. A force-directed algorithm is used to create a discipline map of science in terms of 133 research areas.

A set of significant initiatives in this domain are provided by the Information Visualization Lab (ivl.slis.indiana.edu) and the Cyberinfrastructure for Network Science Center (cns.slis.indiana.edu) at Indiana University directed by Katy Börner, who has focused on creating tools and services for the visualization of knowledge domains (Börner, Chen, & Boyack, 2003; Börner, Sanyal, & Vespignani, 2007).

## The Potential for Science Mapping from Usage Data

As the literature review in the previous section indicates, citation data have held a dominant position in the mapping of science and knowledge domains. However, advances in applications of usage data, for example, social network metrics derived from usage graphs, may translate to the mapping of science and knowledge domains. This approach has not yet found widespread application although some examples have recently emerged in the literature. The general approach is similar to that outlined by Bollen and Van de Sompel (2006b) as shown in Figure 1.18.

Bollen and Van de Sompel (2006b) report on maps created from the flow of usage traffic recorded by the link resolvers of the Los Alamos National Laboratory. These maps are based not on journal or article citation similarities, but journal similarities derived from the flow of usage traffic from one journal to another as users interact with online information services as recorded by an institution's link resolver (Figure 1.19). The maps are based on a principal component analysis of journal usage-similarities that are overlaid with a k-means clustering of journals into particular knowledge domains. The resulting maps reveal the prominent dimensions according to which journals cluster in the online behavior of large groups of users.

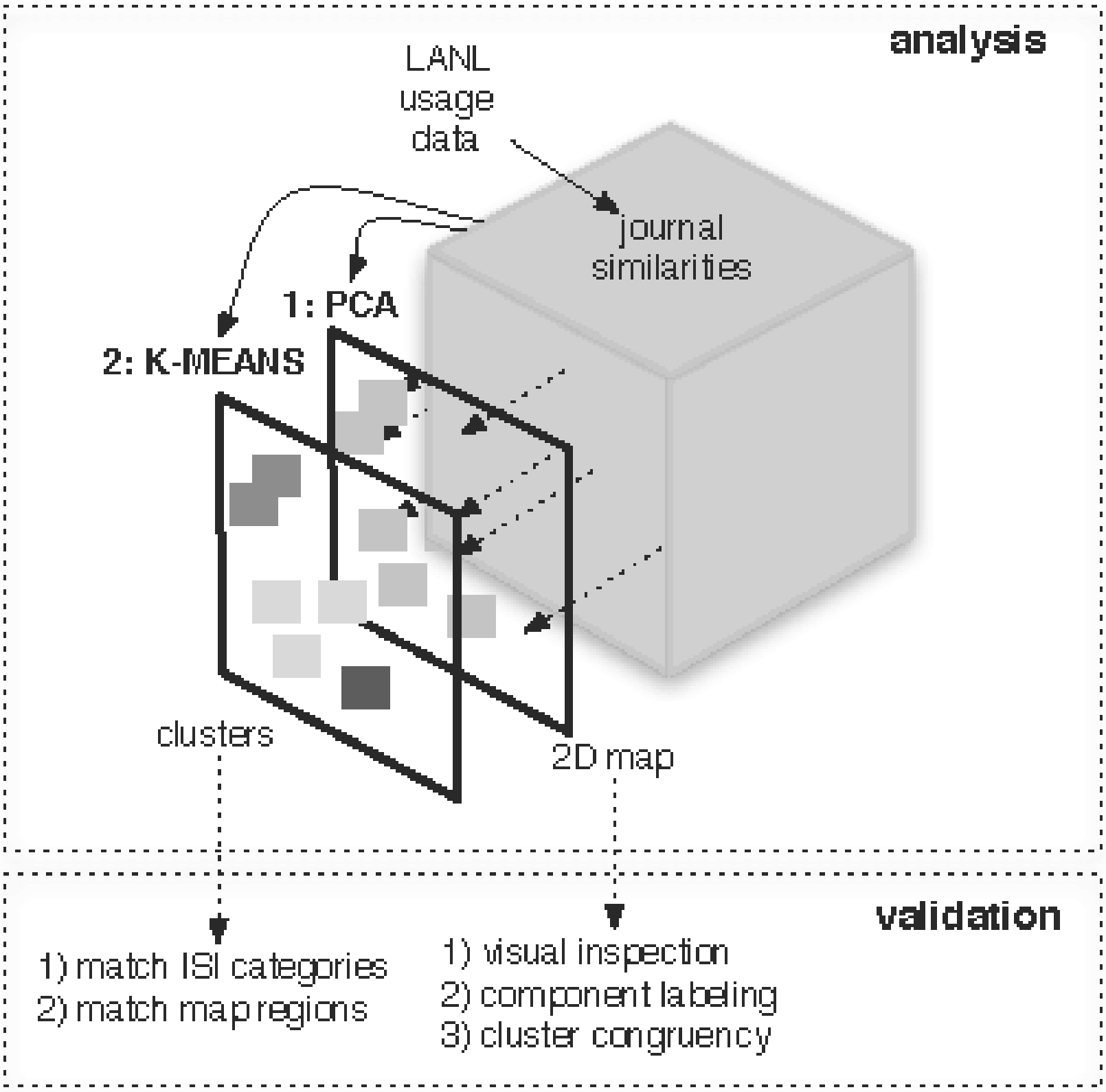

**Figure 1.18 Procedure used to map science from usage (Bollen & Van de Sompel, 2006a)**

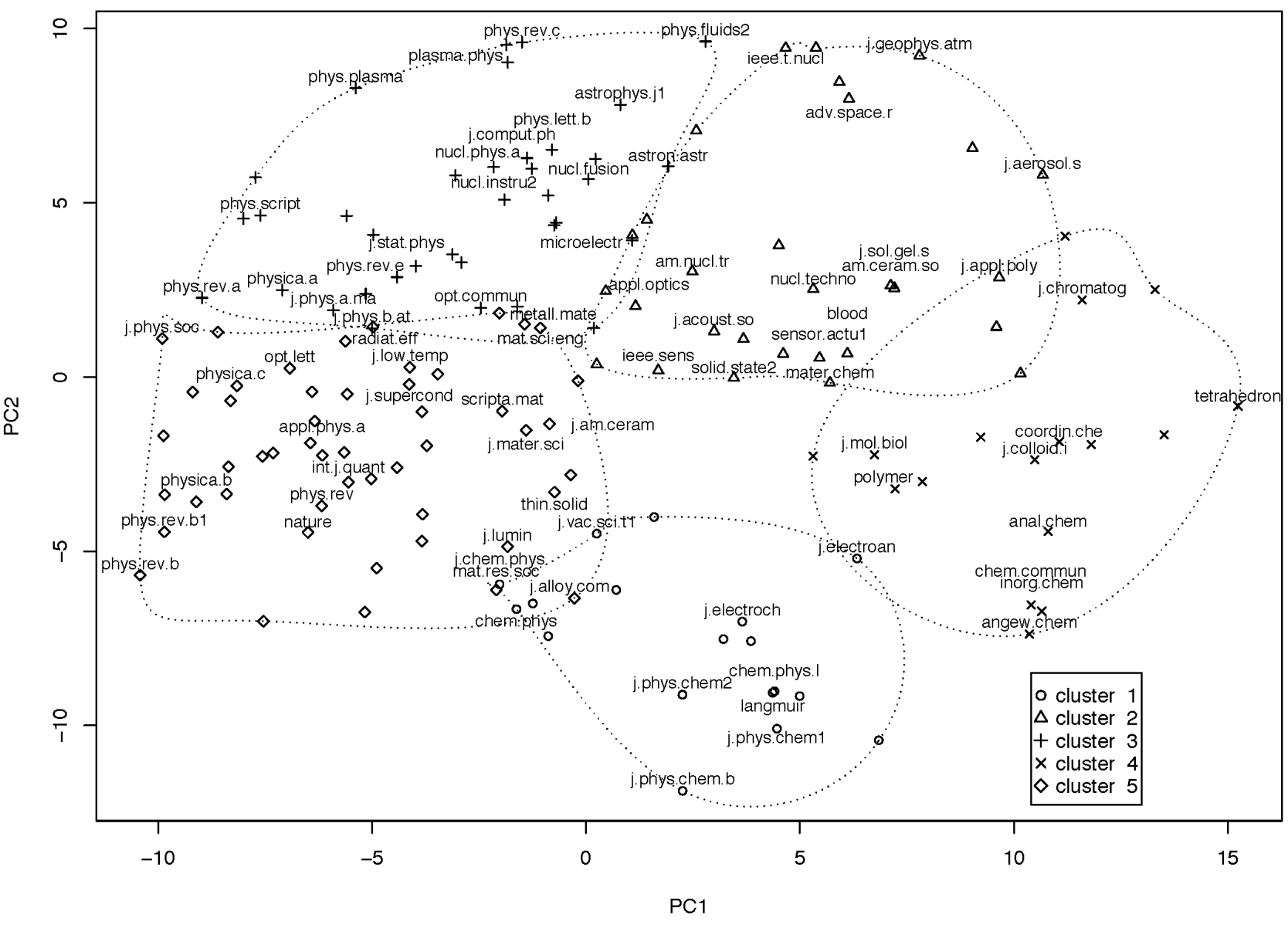

**Figure 1.19 Principal component analysis map of science generated from journal similarities derived from Los Alamos National Laboratory link resolver usage data recorded in 2004. Journal titles are abbreviated to reduce clutter (Bollen & Van de Sompel, 2006a).**

Bollen and colleagues (2008) describe a map of journal clickstreams that was extracted from 200 million usage events that were part of the 1 billion usage events reference data set collected by the MESUR project. This work was later extended with an even larger clickstream data set and a more elaborate validation of the map structure using subject classification taxonomies in Bollen, Van de Sompel, Hagberg, Bettencourt, Chute, Rodriguez, and colleagues (2009a). Figure 1.20 shows a visualization of this network, illustrating the promise of this approach to map the most dominant patterns of traffic in the scholarly community and highlighting the important role of psychology and cognitive science in bridging the social sciences and natural sciences.

Although science mapping from usage has not yet achieved a prominent position in the domain of bibliometrics, the possibilities for tracking science as it takes place are quite promising if certain minimal requirements in terms of sampling and fidelity are met. Sample characterization—whose usage is being mapped—seems to be a crucial issue for attempts to map science and knowledge domains from usage. A particular interest exists in the identification of longitudinal trends in usage data that may inform funding agencies and policy makers of emerging innovation in science.

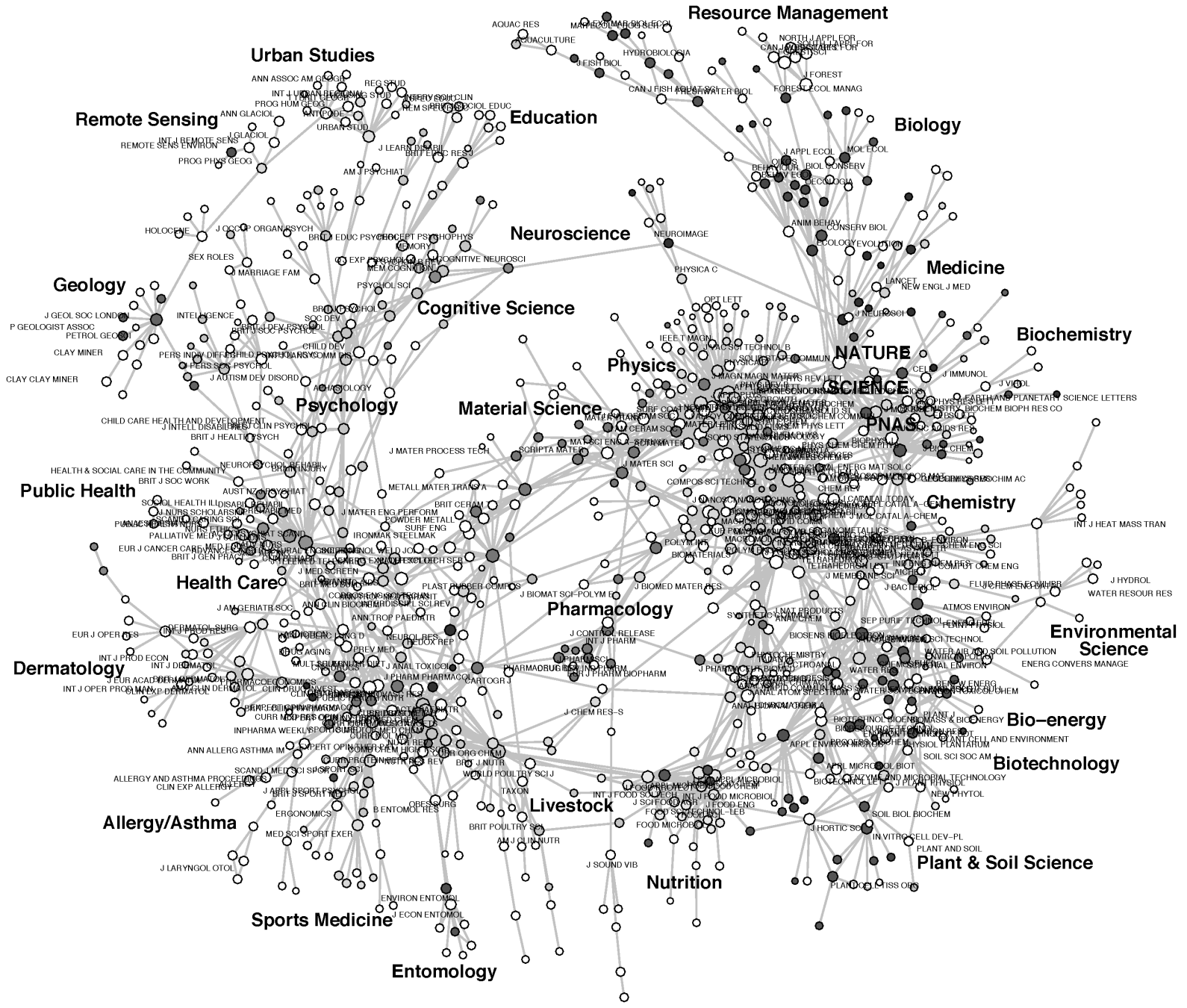

**Figure 1.20 Visualization of journal network extracted from user clickstreams in MESUR's usage data. Journals are represented by circles whose colors correspond to the journal's scientific domain. The lines that connect circles indicate a high probability of one journal following another in users' clickstreams (Bollen et al., 2009a).**

## Conclusion

Bibliometrics is undergoing a renaissance; novel types of data are being combined with powerful new mathematical techniques to create a substantial change. The new techniques are being developed across a wide range of scholarly disciplines, from evolutionary genetics to theoretical physics.

Central to the new bibliometrics is the study of usage and usage patterns. Collections of article level usage event records have existed for only about a decade and their applications are not yet at the level of commonplace acceptance that was long ago reached by article counts and citations. But as scholarly work increasingly moves online, this will change.

Considerable challenges still exist with regard to the standardization of recording and aggregation of usage data. In the present situation usage data are recorded in a plethora of different formats, each representing a different permutation of recording interfaces, data fields, data semantics, and data normalization. The commonalities expressed in the section "A request-based model of usage" can be translated to actual usage data in a variety of different and incompatible ways; this may make it impossible to create usage data sets aggregated across various communities and introduce "lowest-common denominator" limitations. Depending on the data fields that are being recorded and represented, various levels of data loss can take place. For example, privacy concerns can lead to the removal of Internet protocol (IP) addresses from usage data, but without a replacement in the form of an anonymous user ID or session ID, the temporal sequences of user requests are lost. This disables the structural analysis and mapping of the scholarly community, as well as the production of resource rankings such as the social network measures discussed in the section "Social network measures for usage bibliometrics." It is therefore of the utmost importance to consider issues of data loss and fidelity in the development of standards for usage data recording and representation.

Several projects are working to introduce standards to facilitate the recording and aggregation of usage data, for example, COUNTER and MESUR. Standards with regard to the recording and representation of usage data are a sine qua non for usage bibliometrics. However, aggregation of usage data is equally important but often overlooked; it is necessary to arrive at usage bibliometrics that can either exploit or negate the effects of sampling bias. The latter has been shown to greatly affect attempts to produce generalizable impact assessments and perform studies of the scholarly community on the basis of usage data. Bollen and Van de Sompel (2006b) propose an architecture that standardizes the recording, formatting, and aggregation of usage data across various resources and institutions. The proposed architecture presumes the existence of a trusted, third party that aggregates usage data across various participating providers and extracts useful services, for example, ranking and mapping, from the resulting, aggregated data set.

Privacy issues are ubiquitous in usage bibliometrics because the data can reveal the identity and behavior of individual users. However, privacy and confidentiality issues are relevant at three levels, namely that of individual users, their institutions, and the providers (or recorders) of usage data. Different privacy concerns can occur at each level. The protection of the privacy of individual users is not generally a matter of individually negotiated contracts (although stark differences exist in the treatment of individual privacy rights between the U.S. and the EU) but blanket measures can be employed to obscure the identities of users or prevent usage patterns from revealing a user's identity at the point of recording. The use of anonymous session identifiers has been advocated to maintain both temporal information on unique, individual usage patterns as well as user privacy. The protection of confidentiality and privacy at the institutional and provider levels remains within the realm of ad hoc crafted agreements. Usage data ownership presents a challenge as well. Overall the protection of user, institution, and provider privacy stands to benefit from well considered standards for the recording, representation, and aggregation of usage data.

The growth of electronic usage data is but one aspect of the rapid and profound changes affecting the scholarly communication process. For bibliometrics to maintain its traditional role of measuring and analyzing scholarly output, it must adapt and grow with these changes.

Many of the new forms of communication do not lend themselves easily to citation-based analysis. An example is the Sky Server system (Szalay & Gray, 2001, 2006), which provides access to billions of measurements from the Sloan Digital Sky Survey (York et al., 2000). The few thousand citations to the papers describing the dataset do not approach the information contained in the tens of millions of usage events. Singh, Gray, Thakar, Szalay, Raddick, Boroski, and colleagues (2007) have begun to investigate what information can be gleaned from these records. Bollen and colleagues (2008) discuss the discrepancy in scale between citation data and usage data sets by noting that the Web of Science database contains about 600 million article-to-article citations whereas Elsevier alone announced 1 billion full-text downloads in 2006. In other words, aggregations of usage data, such as that constructed by the MESUR project, can easily surpass the total number of citations in existence and thereby significantly enhance the reliability and span of bibliometric studies.

Printed articles are static, but Web services (such as the SDSS Sky Server) are dynamic. In addition to being directly used by individuals, they can be incorporated as part of a different Web service, which in turn can be part of another, and so on. Thus a scholarly Web service can be seen as a node in a complex intellectual infrastructure network. Evaluating the various nodes and links in such a network presents a great challenge for the future of information science. Managing sections of these networks is the task of so-called work flow or provenance systems (Freire, Silva, Callahan, Santos, Scheidegger, & Vo, 2006; Georgakopoulos, Hornick, & Sheth, 1995; Ludascher & Goble, 2005); the logs and audit trails of these systems will provide data to the information scientist, similar to the sets of usage events in today's access logs. Another issue is the meaningful grouping of online resources addressed by the OAI ORE project (www.openarchives.org/ore).

Considerable logistic challenges also lie ahead in applying usage data to scholarly assessment and the analysis of the structure and evolution of the scholarly landscape. Whereas institutional usage data are highly useful for institutional applications, for example, collections management and other services, aggregated usage data sets allow the creation of services and applications that are more generally applicable. Basic compilations, such as the Thomson Reuters' Essential Science Indicators for citation and article count data, can be created from usage data sets that are aggregated across a wide variety of representative institutions. It is an open issue how a representative sample of usage across the entire scholarly community can be achieved, lacking any reliable census data against which any sample can be validated. In addition, such compilations need to be created and maintained in a reliable manner that ensures continuity for extended time periods by parties that are generally trusted to act on behalf of the common good with due respect for privacy, confidentiality, and ownership. Various models can be adopted to support such efforts. However, an environment in which the resulting usage data sets and resulting analysis are least encumbered by rights and ownership issues would be most conducive to the scientific and pragmatic development of this field.

Given the widespread acceptance of *citation*-based measures of impact, such as for the Impact Factor, the final question after all scientific and logistic issues have been addressed will be the community acceptance of *usage*-based measures of article or journal impact. Citation-based measures are seemingly well understood and simple in their applications. Can we expect the same for usage-based measures? A better understanding of sampling issues (whom do the numbers represent?), the nature of usage-based measures of impact (what do the measures mean?), and how and where can they best be applied (what community do they best represent?) will be crucial in establishing usage bibliometrics as a viable complement to citation-based bibliometrics. However, an Albert Einstein quote may best sum up the present situation: "Make everything as simple as possible, but not simpler" (Shapiro, 2006, p. 231). The trade-off to achieve a more comprehensive and accurate usage bibliometrics may very well be one in which widely accepted, simple citation measures are replaced with more complex, but more accurate and reliable usage-based measures.

## Acknowledgments

Michael Kurtz would like to acknowledge the long term collaboration with the Astrophysics Data System (ADS) group at the Smithsonian Astrophysical Observatory, especially Guenther Eichhorn, Alberto Accomazzi, and Edwin Henneken. The ADS is supported by NASA grant NCC5-189. Johan Bollen would like to thank Herbert Van de Sompel and the Digital Library Research and Prototyping Team at the Research Library of the Los Alamos National Laboratory for their support in developing many of the ideas and much of the research presented in this paper. The Andrew W. Mellon Foundation supported the MESUR project to develop usage-based impact metrics from 2006 to 2008.

## Endnote

1. *The Astrophysical Journal* and *Astronomy and Astrophysics* became fully available online on January 1, 1997; *The Monthly Notices of the Royal Astronomical Society* and *The Astronomical Journal* followed on January 1, 1998, although test versions were available earlier.